%% file: rain.tex
\documentclass[12pt]{article}
\usepackage{fullpage,setspace}
\usepackage{amssymb, amsthm, amsmath}
\usepackage{graphicx}
\usepackage[authoryear]{natbib}
\usepackage{bm}
\usepackage{lineno}
\usepackage{comment}
\usepackage{caption}
\usepackage{subcaption}
\usepackage{mathtools}
\usepackage{multirow}
\usepackage{gensymb}
\usepackage[title]{appendix}

\usepackage[T1]{fontenc}
\usepackage[sc]{mathpazo}

\usepackage{algorithm}
\usepackage{algpseudocode}
\usepackage{hyperref}
\usepackage{booktabs}

\input{commands}

\begin{document} % \linenumbers

\begin{center}
{\Large Stochastic Precipitation Generation for the Chesapeake Bay Watershed using Hidden Markov Models with Variational Bayes Parameter Estimation}\\\vspace{6pt}
{\large Reetam Majumder\footnote[1]{North Carolina State University}, Nagaraj K. Neerchal\footnote[2]{University of Maryland, Baltimore County} and Amita Mehta$^2$}\\
\today
\end{center}

\begin{abstract}\begin{singlespace}\noindent
 Stochastic precipitation generators (SPGs) are a class of statistical models which generate synthetic data that can simulate dry and wet rainfall stretches for long durations. Generated precipitation time series data are used in climate projections, impact assessment of extreme weather events, and water resource and agricultural management. We construct an SPG for daily precipitation data that is specified as a semi-continuous distribution at every location, with a point mass at zero for no precipitation and a mixture of two exponential distributions for positive precipitation. Our generators are obtained as hidden Markov models (HMMs) where the underlying climate conditions form the states. We fit a 3-state HMM to daily precipitation data for the Chesapeake Bay watershed in the Eastern coast of the USA for the wet season months of July to September from 2000--2019. Data is obtained from the GPM-IMERG remote sensing dataset, and existing work on variational HMMs is extended to incorporate semi-continuous emission distributions. In light of the high spatial dimension of the data, a stochastic optimization implementation allows for computational speedup. The most likely sequence of underlying states is estimated using the Viterbi algorithm, and we identify the differences in the weather regimes associated with the states of the proposed model. Synthetic data generated from the HMM can reproduce monthly precipitation statistics as well as spatial dependency present in the historical GPM-IMERG data.
 
\vspace{12pt}
\end{singlespace}\end{abstract}

\newpage
\singlespacing

\section{Introduction}
Precipitation is the  major component of the global water cycle and plays an important role in atmospheric and land surface processes in the climate system. While numerical weather models study precipitation over regional to global scales, observational data are used to develop statistical models for precipitation over watershed to local areas at higher temporal frequencies and higher spatial resolution. The measurement and modeling of precipitation has historically relied on sparsely located rain gauge measurements that are spatially irregular. In recent years, precipitation derived from remote sensing observations with uniform spatial and temporal coverage are becoming more easily available. A common class of statistical models which are of interest for analyzing meteorological data are known as stochastic weather generators (SWGs). SWGs can be used to generate multi-year series of synthetic data to simulate weather patterns and are useful in weather and climate research; for precipitation data, the corresponding generator is a stochastic precipitation generator (SPG). The modeling and forecasting of seasonal and inter-annual variations in precipitation is used to determine water allocation and resource management for regions dependent on precipitation as a primary water source. To this end, SPGs produce time series of synthetic data representative of the general rainfall patterns within a region. In particular, they aim to replicate key statistical properties of the historical data like dry and wet stretches, spatial correlations, and extreme weather events. SPGs are also used to downscale precipitation data from numerical weather models, and simulations from them are used for climate projections, impact assessments of extreme weather events, water resources and agricultural management, and for public and veterinary health. Numerical models used in weather and climate research tend to be sensitive to initial conditions, and can be augmented by SWGs. The output from SWGs are stochastic by nature and therefore have uncertainty built in, and ensemble datasets generated from these models can improve other climate and weather models. \citet{brienletal2017} provides a review of current SPG approaches and applications. 

Like most meteorological data, precipitation is modeled as a multivariate time series whose univariate components each correspond to a location. However, modeling it directly using time series methodology usually requires the estimation of a large number of parameters and high-dimensional autocovariance matrices. Daily precipitation data is often modeled as a mixture of a point mass at 0 for no rainfall and one or more Gamma or exponential distributions for positive rainfall \citep{HughesGuttorp1994,Wilks1998,RobertsonKirshnerSmyth2004,MhannaBauwens2012}, introducing an additional layer of complexity. The statistical analysis of such datasets at scale calls for parameter estimation approaches that are computationally efficient while being able to represent the dynamics of the underlying processes to a satisfactory degree. Hidden Markov models (HMMs), initially introduced and studied since the late 1960s \citep{Rabiner1989ATO,Cappeetal2005}, are an attractive class of models that have seen widespread use for constructing SPGs. A hidden Markov model (HMM) is a pair of stochastic processes $\{S_t, Y_t\}_{t \geq 1}$ where $\{S_t\}$ is a Markov chain, and conditional on it, $\{Y_t\}$ is a sequence of independent random variables such that the distribution of $Y_t$ depends only on $S_t$. $\{S_t\}$ usually takes values in a finite set; $t$ is often, although not necessarily, an integer index. However, $\{S_t\}$ is unobservable, and instead we observe only $\{Y_t\}_{t \geq 0}$. $\{Y_t\}$ can be univariate or multivariate, and can follow a discrete, continuous, or mixture distribution. $\{S_t\}$ is known as the state process, while $\{Y_t\}$ is called the emission or observation process. The Markov property of the state process serves to capture the temporal dependency in the data, and the emission process at each time point describes the spatial patterns in the data. Much of the groundwork for using HMMs for daily precipitation was laid in \citet{HughesGuttorp1994}, with \citet{bellone2000} proposing different emission distributions for precipitation amounts and precipitation occurrence models. This was extended to non-homogeneous hidden Markov models by \citep{RobertsonKirshnerSmyth2004, RobertsonEtAl2006, KirshnerThesis}, where the transition probabilities of the HMM's Markov process change over time. 

The overwhelming majority of HMM studies use the Baum-Welch algorithm \citep{baum1966,baum1967,baum1968,baum1970,Baum1972} for parameter estimation. It is a maximum likelihood approach used for efficient parameter estimation in HMMs while taking into account the Markov assumptions of the model, and can be considered as a variant of the expectation-maximization (EM) algorithm \citep{DempsterElAl1977}. The Viterbi algorithm \citep{Viterbi} can then estimate the most likely sequence of states that has generated the data. The ability to estimate and interpret the underlying states of a relatively parsimonious model has made HMMs a popular approach for sequential data.
  However, the Baum-Welch algorithm, being a maximum likelihood based method, can run into problems for large datasets with complex dependencies. In particular, it can lead to model overfitting for graphical models which tend to have complex dependency structures  \citep{attias1999}. \citet{Holsclaw2016} use a Bayesian approach to model daily precipitation, but in general, Bayesian alternatives which use Gibbs sampling \citep{scott2002,Cappeetal2005} tend to be computationally intensive. Historically, the reliance on spatially non-uniform weather stations for data has prevented these from being practical issues. However, as gridded remote sensing data which tend to be highly correlated become more easily available, alternative approaches which are scalable and can incorporate prior information are desirable.
 This is where variational Bayes (VB) provides an attractive alternative for parameter estimation. While Markov chain Monte Carlo (MCMC) methods use sampling to find the posterior distribution, VB uses optimization to calculate an approximate posterior; the posteriors are obtained by an iterative EM-like algorithm which always converges \citep{attias1999}. The variational posteriors have analytical forms under certain conditions \citep{Grahramanietal} and can be used to perform approximate Bayesian inference. A review of VB methods can be found in \citet{bleietal2017}. However, while VB estimation has been implemented for state space models and HMMs \citep{MacKay97ensemblelearning,Grahramanietal,Beal03variationalalgorithms,Ji2006VariationalBF,mcgroryetal}, studies have usually only focused on cases where emissions are distributed as Normal or mixtures of Normal distributions. 
 
In this paper, we outline VB estimation for HMMs with semi-continuous emissions, with the motivation of constructing an SPG for daily precipitation using gridded remote sensing data from GPM-IMERG \citep{IMERG} for a large spatial domain. Our model is constructed using precipitation data for the Chesapeake Bay watershed in Eastern US for the wet season months of July--September of 2000--2019. 
%Restricting ourselves to a season allows us to keep the Markov chain parameters constant over time, i.e., a homogeneous HMM. 
The SPG aims to replicate the spatial correlation present in the data, as well as key properties of the original data, e.g., the proportion of dry days (with no rainfall) and mean seasonal rainfall. Estimates for these can be calculated using data simulated from the fitted model.

The rest of this paper is organized as follows: Section 2 provides background for HMMs and VB. Section 3 introduces the dataset and discusses the HMM for precipitation as well as VB estimation for the model. Section 4 presents a numerical study for multi-site precipitation, and also presents our case study for daily precipitation over the Chesapeake Bay watershed. Section 5 concludes with a discussion.
%%%%%%%%%%%%%%%%%%%%%%%%%%%%%%%%%%%%%%%%%%
\section{Background}
We provide some background on parameter estimation for hidden Markov models and on variational Bayes in this section. A more thorough treatment of learning procedures for HMMs as well as parameter estimation using variational Bayes can be found in \citet[Chapter 2]{ReetamThesis}.
\subsection{Hidden Markov models}
 \begin{figure}
    \centering
     \includegraphics[width=0.9\linewidth]{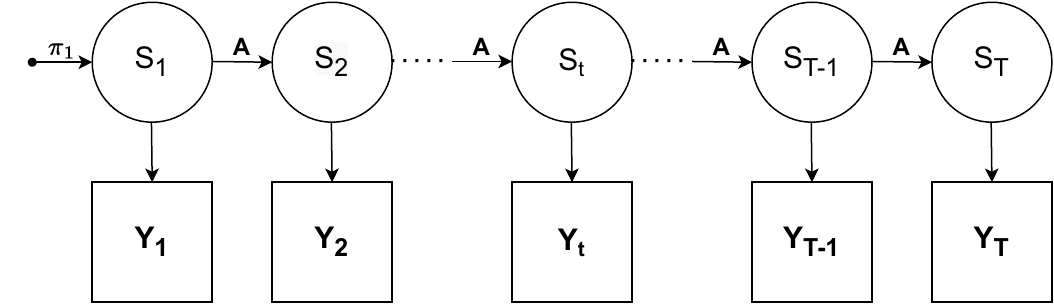}
\caption{ A graphical model representation of the conditional independence structure for an HMM.}
    \label{fig:hmm}
\end{figure}
An HMM consists of a sequence of multivariate observations $\textbf{y}_{1:T} = [\textbf{y}_1, \ldots, \textbf{y}_T]$, together with a sequence of hidden (unobserved) states $s_{1:T} = [s_1,  \ldots, s_T]$. The states are assumed to follow a first order Markov process, and the multivariate observation $\mathbf{y}_t = (y_{t1},\ldots,y_{tL})'$ is emitted by the corresponding state $s_t \in \{1, \ldots , K\}$. For the purposes of this study, $L$ can be considered the number of spatial locations. Figure \ref{fig:hmm} shows a graphical model representation of an HMM. The state process $s$ is 
parameterized by an initial probability $\pi_{1j} = Pr[s_1=j]$ and a $K\times K$ matrix $A$, whose elements are $a_{jk} = Pr[s_{t+1} = k|s_t = j]$ for $j,k = 1,\ldots K$. The probability density of the emission $y_{tl}$ at location $l$ and time $t$, given that the system is in state $j$ , is:
\begin{align*}
    p(y_{tl}|s_t = j) &= p_j(y_{tl}|\theta_{jl}),
\end{align*}
where $\theta_{jl}$ are the parameters associated with the distribution of $y_{tl}$. The distribution at each location are assumed to be independent conditional on the state, and the full likelihood can be expressed as:
\begin{align*}
    p(y,s|\Theta) = \pi_{1j}\prod_{t=1}^{T-1}\prod_{j=1}^K \prod_{k=1}^K a_{jk} \prod_{t=1}^T\prod_{j=1}^K \prod_{l=1}^Lp_j(y_{tl}|s_{tj}),
\end{align*}
where $s_{tj} = \mathbb{I}\{s_t=j\}$. We refer the reader to \citet{Rabiner1989ATO} for a detailed tutorial on parameter estimation using the Baum-Welch algorithm.
\subsection{Variational Bayes inference}
Variational Bayes (VB) methods aim to approximate the posterior distribution through optimization. VB tends to be faster than MCMC for intractable likelihoods, but it only provides approximate inference. VB is suited for large datasets, and can take advantage of stochastic optimization \citep{RobbinsMonro1951} which makes it scalable. VB posits a family of approximate posterior distributions $\mathbb{Q}$ over the latent variables $\boldsymbol z$ and parameters $\theta$, and optimizes within this family to find the member closest to the true posterior $p(\boldsymbol{z},\theta|y)$. In its most widely applied form, the VB posterior minimizes the Kullback-Leibler (KL) divergence \citep{KullbackLeibler1951} to the true posterior among all candidates $q(\cdot) \in \mathbb{Q}$, i.e.,
\begin{equation}\label{eqn:KLVB}
    \Tilde{q}(\cdot) = {\arg \min_{q(\cdot) \in \mathbb{Q}}} KL\bigl(q(\boldsymbol{z},\theta)\parallel  p(\boldsymbol{z},\theta|\boldsymbol{y})\bigr).
\end{equation}
  Optimizing the KL divergence is typically difficult in practice since it involves computing the log marginal likelihood $\log p(\boldsymbol{y})$. 
  %The optimum values of the hyperparameters are found using a variational EM-like algorithm. Note that the true posterior is typically not in the variational family $\mathbb{Q}$.
However, it is possible to find a lower bound for $\log p(\boldsymbol{y})$ and equivalently maximize a quantity known as the evidence lower bound (ELBO) \citep{jordanetal1999}, defined as
\begin{equation}\label{eqn:ELBO}
    \text{ELBO}(q) = \mathbb{E}[\log p(\boldsymbol{z},\theta,\boldsymbol{y})] - \mathbb{E}[\log q(\boldsymbol{z},\theta)].
\end{equation}
The so called mean-field assumption is commonly made to factorize  $q(\boldsymbol{z},\theta)$ by assuming independence between the variational posterior of the parameters and latent variables:
\begin{align}
    q(\boldsymbol{z},\theta) \approx q(\boldsymbol z) q(\theta).
\end{align}
The ELBO is optimized by updating $q(\boldsymbol z)$ and $q(\theta)$ in turn by using a variational Bayes expectation-maximization (VBEM) algorithm. In particular, if the complete data likelihood is in the exponential family and we choose conjugate priors, the VBEM updates have analytical expressions. The mean-field assumption often extends to the components of $q(\boldsymbol z)$ and $q(\theta)$ as well; \citet{Grahramanietal} provide the general forms of the variational updates for these conjugate exponential models under complete mean-field factorization. This is known as mean-field variational Bayes, and optimizing the ELBO one parameter at a time using the entire available data is often referred to as coordinate ascent variational inference (CAVI) \citep{bleietal2017}.

\subsection{Stochastic variational Bayes}
The VBEM algorithm can be bottlenecked by the data length since posterior means for the parameters and latent variables need to be computed at every iteration, requiring a pass through the entire data. Stochastic optimization methods can provide computational speedup, and stochastic variational Bayes (SVB) \citep{JMLR:v14:hoffman13a} is one such approach which modifies the VBEM algorithm into a stochastic gradient ascent algorithm for each parameter. Instead of computing gradients based on the entire data, SVB uses an unbiased estimate of the gradient at each iteration.
Let $\mathcal{L}(\lambda)$ be the ELBO for a parameter $\lambda$ that needs to be maximized using VB optimization, with $\nabla_\lambda \mathcal{L}(\lambda)$ denoting its gradient. Next, let $B(\lambda)$ be a random function that is unbiased estimator of $\nabla_\lambda \mathcal{L}(\lambda)$, i.e. $\mathbb{E}_q B(\lambda) = \nabla_\lambda \mathcal{L}(\lambda)$. For example, $B(\lambda)$ could be the gradient computed from random samples, or minibatches, taken from the entire data. Then the stochastic gradient ascent step for optimizing the ELBO for $\lambda$ is
\begin{align*}
    \lambda^{i} = \lambda^{i-1} + \tau_i\cdot b_i(\lambda^{i-1})
\end{align*}
for step size $\tau_i$, where $b_i(\cdot)$ is an independent draw from the noisy gradient $B(\cdot)$. If $\tau_i$ satisfies the Robbins-Monro conditions \citep{RobbinsMonro1951}, namely
\begin{align*}
    \sum_i \tau_i &= \infty\\
    \mbox{and } \sum_i \tau_i^2 &< \infty,
\end{align*}
then $\lambda^i$ converges to a local optimum. If $G_i$ is any positive definite matrix of appropriate dimensions, a similar gradient ascent property holds \citep{JMLR:v14:hoffman13a}, with parameter update at step $i$ given by:
\begin{align}\label{eq:svi1}
    \lambda^{i} = \lambda^{i-1} + \tau_i\cdot G_i^{-1}b_i(\lambda^{i-1}).
\end{align}
In particular, if we choose $G_i = \mathcal{G}_i$, the Fisher information matrix, the resulting natural gradient provides the direction of the steepest ascent for the optimization. In the mean field setup, the noisy gradient is often obtained by randomly sampling a single data point and doing all computations based on that single data point. 
 \citet{JMLR:v14:hoffman13a} have showed a direct relationship between CAVI updates and SVI updates for models belonging to the conjugate exponential family that we take advantage of in our study.

The main difficulty in implementing SVI for HMMs comes from the dependency of stochastic optimization on samples, or minibatches, from the data. HMMs are time dependent, and thus sequential draws are required if we want to sample from the process. We denote this sequential sample, or minibatch, as $y^*$. The nature of the dataset dictates the procedure for selecting $y^*$. If the data consists of a single long chain, \citet{NIPS2014_Fotietal} proposed subsampling from the chain and buffering the beginning and end with extra observations to preserve the Markov properties of the states. If, however, the data is seasonal or cyclical in nature that can be represented as \textit{N} blocks each of size \textit{D}, 
then a minibatch is constructed at each optimization iteration by randomly sampling blocks with replacement and selecting all $D$ time points within the block. This approach is discussed in \citet{JohnsonWillsky2014}. In both cases, the variational E-step employs the Forward-Backward algorithm, and the variational M-step can take advantage of conjugate priors and provides parameter updates through stochastic gradient ascent.
%%%%%%%%%%%%%%%%%%%%%%%%%%%%%%%%%%%%%%%%%%
\section{Data and Methodology}
\subsection{Remote sensing data from GPM-IMERG}

\begin{figure}
    \centering
    \begin{subfigure}[b]{0.46\linewidth}
     \includegraphics[width=\linewidth]{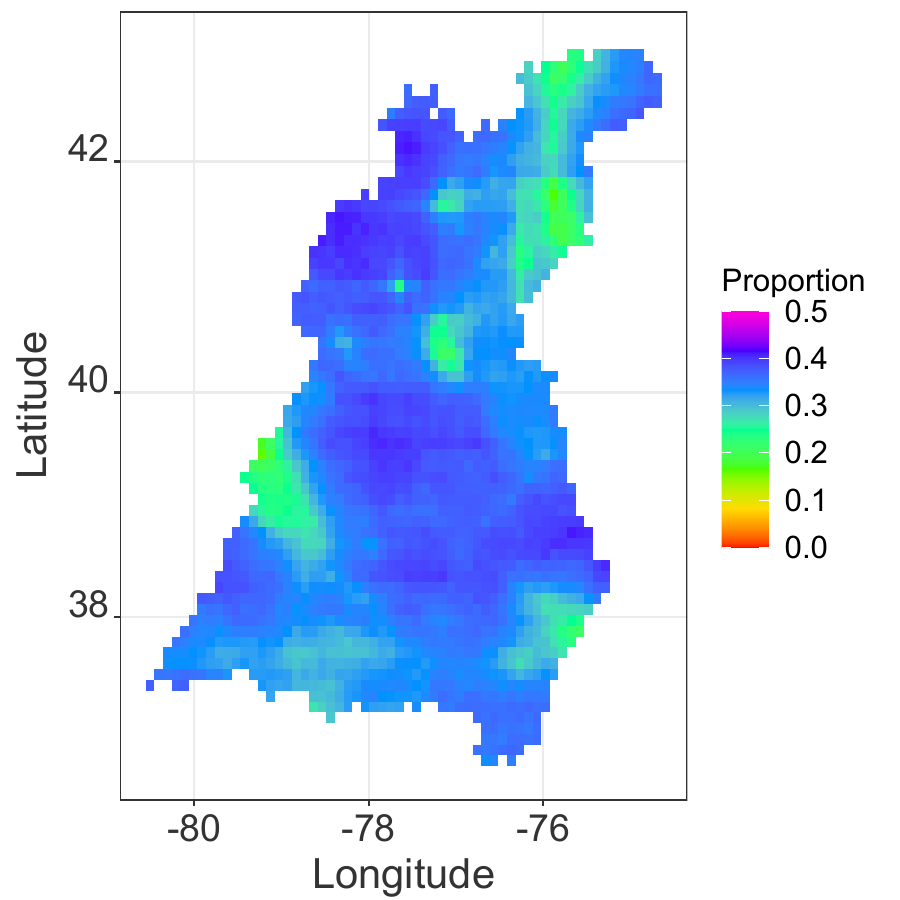}
\caption{Proportion of dry days at each location.}
    \label{fig:cbay_obs_dry}
    \end{subfigure}
    \hfill
    \begin{subfigure}[b]{0.46\linewidth}
        \ \includegraphics[width=\linewidth]{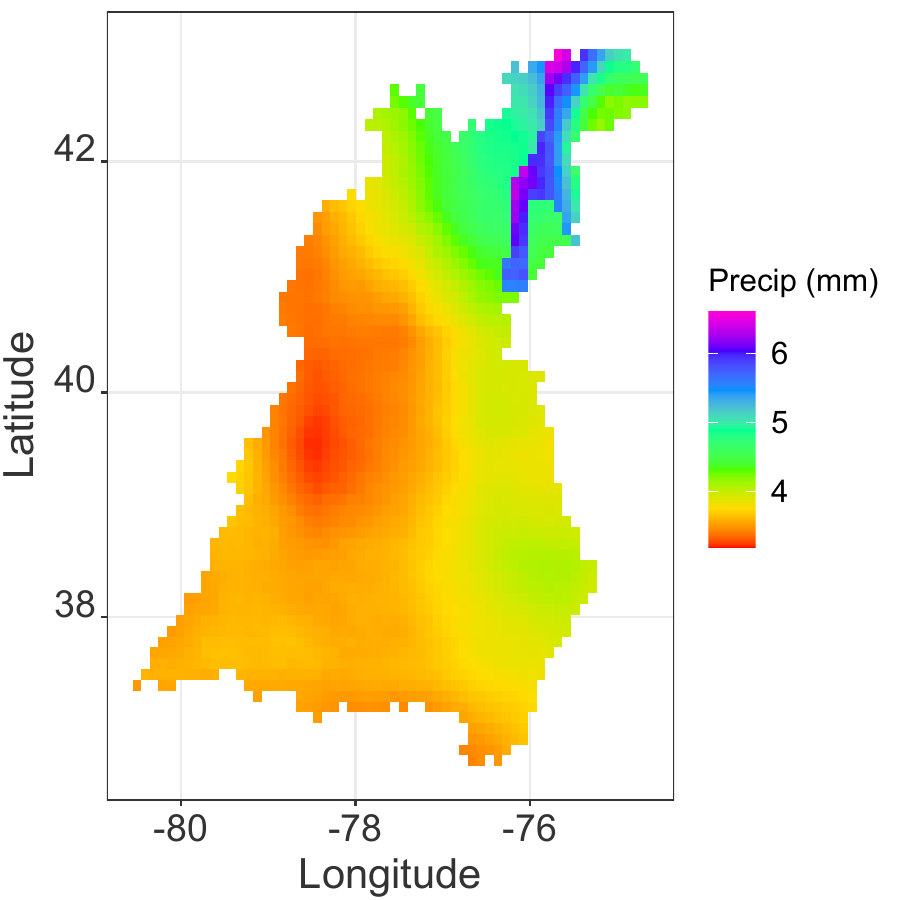}
\caption{Mean daily precipitation at each location.}
    \label{fig:cbay_obs_wet}
    \end{subfigure}
        \caption{Proportion of dry days and mean daily precipitation (mm) during Jul--Sep for the Chesapeake Bay watershed from GPM-IMERG data, averaged over 2000--2019.}
    \label{fig:imerg}
\end{figure}

The Global Precipitation Measurement (GPM) mission %\footnote{\url{https://gpm.nasa.gov/missions/GPM/constellation}} 
is an international satellite mission co-led by the National Aeronautics and Space Administration (NASA) and the Japanese Aerospace and Exploration Agency (JAXA). It aims to combine precipitation measurements from multiple research and operational microwave sensors from a constellation of international satellites for delivering global precipitation products. 
The Integrated Multi-satellitE Retrievals for GPM (IMERG) algorithm\footnote{\url{https://gpm.nasa.gov/data/imerg}} \citep{IMERG} combines the information that is made available through the GPM constellation satellites to provide global precipitation estimates. The GPM-IMERG product provides inter-calibrated precipitation estimates collected during the operation of the Tropical Rainfall Measuring Mission (TRMM) research satellite %\footnote{\url{https://gpm.nasa.gov/missions/trmm}} 
(2000--2015) and more recent precipitation estimates from GPM (2014--present). This provides over 20 years of record of global precipitation that can be used by researchers to better validate climate and weather models and understand long-term trends and extreme precipitation patterns.

Our case study encompasses the Chesapeake Bay watershed in Eastern USA. We use daily data from the GPM-IMERG dataset for the months of July to September from 2000--2019. At a spatial resolution of $0.1\degree \times 0.1\degree$, The IMERG dataset covers the 64,000 square mile watershed with 1927 grid points. Figure \ref{fig:imerg} plots the proportion of dry days and daily mean precipitation averaged over the 20 years, which show a high degrees of spatial variation between the grid points. Our analysis focuses on July--September since they are the wettest months of the year for this area. Figure \ref{fig:cbay_obs_dry} identifies the driest regions of the watershed during the wet season, mostly in the north-west and central areas. Figure \ref{fig:cbay_obs_wet} shows smooth precipitation gradients over most of the watershed. Overall, there are a wide range of precipitation values observed in the data. The HMM aims to identify the underlying precipitation patterns and be able to generate synthetic precipitation from the model which can replicate the empirical behavior of the data.

 \subsection{The HMM for precipitation}\label{s:precipHMM}
For daily precipitation data at $L$ locations, we consider $y_{tl}$ to be distributed as an $M+1$ component mixture with a point mass at 0 for no rainfall, and $M$ exponential distributions for positive rainfall. For each state $j$, we define the indicator variable $r_{tjlm}$ to connect the underlying state to the emission distribution such that:
\begin{align*}
    r_{tjlm}s_{tj} = &\mathbb{I}\{y_{tl} \text{ comes from the } m^{th} \text{ mixture component and } s_t=j\},
\end{align*}
with $m=0,1,\ldots,M$, and $l = 1,\ldots, L$. The $m=0$ mixture component corresponds to the point mass. The number of states (\textit{K}), the number of locations (\textit{L}), and the number of mixture components (\textit{M}+1) in the HMM are assumed to be known.
For each state \textit{j} and location \textit{l}, $r_{tjl} = (r_{tjl0}, \ldots, r_{tjlm})$ follows a categorical distribution:
\begin{align}
    p(r_{tjl}|c_{jl},s_t=j) &= \prod_{m=0}^M c_{jlm}^{r_{tjlm}},
    %\\
  % \mbox{and\hspace{5pt} } p(r_{tjl}|c_{jl},s_t = j) &\perp \!\!\! \perp p(r_{tjl'}|c_{jl'},s_t=j) \mbox{\hspace{5pt} for } l \neq l', \label{eq:mix}
\end{align}
with $m=0,1,\ldots,M$ and $l = 1,\ldots, L$; 
%$c_{jl} = (c_{jl0},\ldots, c_{jlM})$ are the mixture probabilities parameterizing $r_{tjl}$, with 
$c_{jlm}\geq 0 $ for all \textit{m}, and $\sum_{m=0}^M c_{jlm}=1 $. 
%A consequence of the conditional independence assumption in \eqref{eq:mix} is that the correlation between precipitation at different locations is induced by the common state variable for all locations. 
If we assume that positive rainfall at the $l^{th}$ location from the $m^{th}$ mixture component (where  $m>0$) arising from state \textit{j} follows an exponential distribution with rate $\lambda_{jlm}$, the distribution of an observation from state \textit{j} across all locations is given by
\begin{align}
\begin{split}
       \prod_{l=1}^L p_j(y_{tl},r_{tjl}|\lambda_{jl},c_{jl},s_t=j) &= \prod_{l=1}^L p(r_{tjl}|c_{jl},s_t=j) \cdot p(y_{tl}|\lambda_{jl},r_{tjl},s_t=j)\\ &=\prod_{l=1}^L \biggl\{c_{jl0}^{r_{tjl0}}\prod_{m=1}^M\bigl[c_{jlm}\lambda_{jlm}\exp\{-\lambda_{jlm}y_{tl}\}\bigr]^{r_{tjlm}}\biggr\}.
\end{split}
\end{align}
Note that the distributions of rainfall at different locations are mutually dependent through the underlying state process $s$. We denote $C_l=((c_{jlm}))$ as the $K\times (M+1)$ matrix of mixture probabilities for each location \textit{l}, with $C = (C_1, \ldots, C_L)$. Similarly, $\Lambda_l = ((\lambda_{jlm}))$ is a $K\times M$ matrix whose elements are the independently distributed rate parameters of the exponential distributions, with $\Lambda = (\Lambda_1, \ldots, \Lambda_L)$. The complete data likelihood can be expressed as:
\begin{equation}\label{eqn:condLikExp2}
\begin{split}
            p(y,s,r|\Theta) &= \prod_{j=1}^K \bigl \{\pi_{1j} \bigr\}^{s_{1j}} \prod_{t=1}^T\prod_{j=1}^K \prod_{l=1}^L \bigl \{p_j(y_{tl},r_{tjl}|\Theta)\bigr\}^{s_{tj}}
            \prod_{t=1}^{T-1}\prod_{j=1}^K\prod_{k=1}^K\bigl\{a_{jk}\bigr\}^{s_{tj}s_{t+1,k}} \\
             &= \text{ } \exp\biggl\{\sum_{j=1}^K s_{1j}\log \pi_{1j} + \sum_{t=1}^{T}\sum_{j=1}^K \sum_{l=1}^L \biggl[\sum_{m=1}^M s_{tj}r_{tjlm}(\log c_{jlm} + \log \lambda_{jlm} - y_t\lambda_{jlm}) \\
         &\text{ \hspace{5 mm}} + s_{tj}r_{tjl0}\log c_{jl0}\biggr] + \sum_{t=1}^{T-1} \sum_{j=1}^K \sum_{k=1}^K s_{tj}s_{t+1,k}\log a_{jk} \biggr\}.
         \end{split}
\end{equation}

Similarly, we write the prior as:

\begin{align}\label{eqn:prior}
\begin{split}
    p(\Theta|\nu^{(0)}) &= p(\pi_1)\cdot p(\lambda)\cdot p(C)\cdot p(A)\\
    &=\exp\biggl\{\sum_{j=1}^K\bigl\{(\xi^{(0)}_j - 1)\log \pi_{1j} + \sum_{l=1}^L \sum_{m=1}^M \bigl[ -\delta_{jlm}^{(0)}\lambda_{jlm} + (\gamma_{jlm}^{(0)}-1)\log \lambda_{jlm}\bigr]\\ 
    &\text{\hspace{25pt}} + \sum_{l=1}^L(\zeta_{jl0}^{(0)}-1)\log c_{jl0} + \sum_{l=1}^L \sum_{m=1}^M (\zeta_{jlm}^{(0)}-1)\log c_{jlm}\\
    &\text{\hspace{50pt}} + \sum_{k=1}^K (\alpha_{jk}^{(0)}-1)\log a_{jk}\bigr\} - \log h^{(0)}\biggr\},
\end{split}
\end{align}
where $h^{(0)} = h(\nu^{(0)})$ is the normalizing constant for the prior.
There are $\mathcal{O}(KLM)$ parameters in the model. For the Chesapeake Bay watershed, $L=1927$ and $M=2$, whereas the number of data points $T=1840$. In general, the number of parameters will always exceed the data size for even moderately large spatial problems. Bayesian approaches are especially useful in this context since they can add prior information which also makes the model identifiable.

\subsection{VB parameter estimation for the HMM}

Parameter estimation in HMMs employ the Baum-Welch (BW) algorithm \citep{baum1970}, which can be expressed as a special case of the EM algorithm. The primary difference is in the E-step, where recursive forward and backward algorithms maintain the Markov nature of the state process. The BW algorithm translates seamlessly to a variational context. Since the distribution of the state process in HMMs does not factorize completely due to its Markov property, this can be considered part of a more general class of variational procedures called structured variational Bayes which partially relaxes the mean-field assumption. \citet{Ji2006VariationalBF} describes the VBEM algorithm for HMMs where the emissions are mixtures of continuous distributions. However, there is not, to the best of our knowledge, studies exploring semi-continuous mixture models and how their posteriors are affected by the variational approximation. 

For our HMM for precipitation, the complete data likelihood is given by:
\begin{align}
    p(y,s,r|\Theta) &= p(y,r|s,\Theta)\cdot p(s|\Theta),\nonumber
\end{align}
where $\Theta = (A, C, \Lambda, \pi_1)$ parameterizes the HMM. The variational family $\mathbb{Q}$ for the posterior is constrained to distributions which are separable in the following manner:
\begin{align}
    q(\Theta,s,r) &= q_\Theta(\Theta)\cdot q_{s,r}(s,r), \label{eqn:varposteriorexp2} \\
    \text{where }q_\Theta(\Theta) &= q(\pi_1)\cdot q(A)\cdot q(C)\cdot q(\Lambda) \label{eqn:qthetaexp2}.
\end{align}
We assume the prior decomposes as:
\begin{align*}
    p(\Theta|\nu^{(0)}) &= p(\pi_1)\cdot p(A)\cdot p(C)\cdot p(\Lambda),
\end{align*}
where $\nu^{(0)}$ are known hyperparameters. The individual components of the prior are distributed as:
\begingroup
\allowdisplaybreaks
\begin{align*}
    p(\pi_1) &= \mbox{Dirichlet}(\pi_1|\xi^{(0)}),\\
    p(A) &= \prod_{j=1}^K \mbox{Dirichlet}(a_j|\alpha_j^{(0)}), \\
    p(C) &= \prod_{j=1}^K \prod_{l=1}^L  \mbox{Dirichlet}(c_{jl}|\zeta_{jl}^{(0)}), \\
   \text{and } p(\Lambda) &= \prod_{j=1}^K \prod_{l=1}^L \prod_{m=1}^M \mbox{Gamma}(\lambda_{jlm}|\gamma_{jlm}^{(0)}, \delta_{jlm}^{(0)}),
\end{align*}
where $a_j = (a_{j1},\ldots,a_{jK})$,  $\pi_1 = (\pi_{11}, \ldots, \pi_{1K})$, $\zeta_{jl}^{(0)} = ( \zeta_ {jl0}^{(0)},\ldots, \zeta _{jlM}^{(0)})$, $\alpha_j^{(0)} = (\alpha_{j1}^{(0)},\ldots, \alpha_{jK}^{(0)})$, and $\xi^{(0)} = (\xi_1^{(0)},\ldots\xi_K^{(0})$. $\gamma_{jlm}^{(0)}$ and $\delta_{jlm}^{(0)}$ are the shape and rate parameters of the Gamma distribution respectively. 

The VBEM algorithm iterates between updating the posterior for the model parameters, $q_\Theta(\Theta)$, and the posterior for the latent variables, $q_{s,r}(s,r)$. Posterior updates of $q_{\Theta}(\Theta)$ and $q_{s,r}(s,r)$ can be obtained as closed form expressions since our model is in the conjugate exponential family \citep{Grahramanietal}. The M-step (VBM) and E-step (VBE) of the VBEM algorithm are outlined below.

\noindent \textbf{VBM step}: \textit{Fix $q_{s,r}(s,r)$ at its expected values and update $q_\Theta(\Theta)$.}

Since  $q_\Theta(\Theta)$ is conjugate to the prior, its posterior distribution is obtained by updating the coordinates of $\nu^{(0)}$ with the expected values of the corresponding sufficient statistics $u(s,y,r)$. To this end, we denote the expectations of the latent variables in \eqref{eqn:condLikExp2} under $q_{s,r}(s,r)$ as
\begingroup
\allowdisplaybreaks
\begin{align*}
    q_{1j} &= \mathbb{E}(s_{1j}),\\
    q_{tj} &= \mathbb{E}(s_{tj}),\\
    q_{tjlm} &= \mathbb{E}(r_{tjlm}),\\
    \text{and }q_{jk} &= \mathbb{E}(s_{tj}s_{t+1,k}),
\end{align*}
where $j,k=1,\ldots,K$, $l=1,\ldots, L$, and $m=0,1,\ldots,M$. The variational updates at each iteration of the VBM step are then given by
\begin{align*}
        \xi_j &\leftarrow \xi_j^{(0)} + q_{1j},\\
        \zeta_{jl0} &\leftarrow \zeta_{jl0}^{(0)} + \sum_{t=1}^T q_{tj}q_{tjl0}, \\
\zeta_{jlm} &\leftarrow \zeta_{jlm}^{(0)} + \sum_{t=1}^T q_{tj}q_{tjlm}, \\
\gamma_{jlm} &\leftarrow \gamma_{jlm}^{(0)} + \sum_{t=1}^Tq_{tj}q_{tjlm}, \\
\delta_{jlm} &\leftarrow \delta_{jlm}^{(0)} + \sum_{t=1}^Tq_{tj}q_{tjlm}y_{tl}, \\
\alpha_{jk} &\leftarrow \alpha_{jk}^{(0)} + \sum_{t=1}^{T-1} q_{jk} ,
    \end{align*}
    where $j,k=1,\ldots,K$, $l=1, \ldots, L$, and $m=1,\ldots,M$.
\endgroup

\noindent \textbf{VBE step}: \textit{Fix $q_\Theta(\Theta)$ at its expected values and update $q_{s,r}(s,r)$.}

The variational posterior $q_{s,r}(s,r)$ has a form similar to the complete data likelihood, i.e.,
\begingroup
\allowdisplaybreaks
\begin{align}
    q_{s,r}(s,r) &\propto \prod_{j=1}^K \bigl\{{a_{1j}^*} \bigr\}^{s_{1j}}\prod_{t=1}^T \prod_{j=1}^K \prod_{l=1}^L \prod_{m=0}^M \bigl\{ {b_{tjlm}^*}\bigr\} ^{s_{tj}r_{tjlm}} \prod_{t=1}^{T-1} \prod_{j=1}^K \prod_{k=1}^K \bigl\{ {a_{jk}^*}\bigr\}^{s_{tj}s_{t+1,k}} \label{eqn:qsr},
    \end{align}
    with the natural parameters $\phi(\Theta)$ replaced by their expectations under $q_\Theta(\Theta)$. Comparing with \eqref{eqn:condLikExp2}, we get
    \begin{align*}
    a_{1j}^* &= \exp\bigl\{\mathbb{E}_q \log \pi_{1j} \bigr \} = \exp\bigl\{\Psi(\xi_j) - \Psi(\xi_{.})\bigr\},  \\
    \text{and \hspace{5pt}} a_{jk}^* &= \exp\bigl\{\mathbb{E}_q \log a_{jk} \bigr \} = \exp\bigl\{\Psi(\alpha_{jk}) - \Psi(\alpha_{j.})\bigr\},
    \end{align*}
    where  $ \xi_\cdot = \sum_{j=1}^K \xi_j$ , $\alpha_{j.} = \sum_{k=1}^K \alpha_{jk}$. Similarly,
    \begin{align*}
    b_{tjlm}^* &= \begin{cases} \exp\bigl\{\mathbb{E}_q \log \bigl [ c_{jl0} \bigr]\bigr \}&\mbox{ if } m=0, \nonumber\\
    %\\
    \exp\bigl\{\mathbb{E}_q \log \bigl [ c_{jlm}f(y_{tl}|\lambda_{jlm}) \bigr]\bigr \} &\mbox{ if } m>0.
    \end{cases}
     \end{align*}
      The expectations of the individual terms in $b_{tjlm}^*$ are:
   \begin{align*}
    c_{jlm}^* &= \exp\bigl\{\mathbb{E}_q \log c_{jlm}\bigr\} = \exp\bigl\{\Psi(\zeta_{jlm}) - \Psi(\zeta_{j.})\bigr\}, 
    \text{ where } \zeta_{j.} = \sum_{m=0}^M \zeta_{jlm}, \\
    \lambda_{jlm}^* &= \exp\bigl\{\mathbb{E}_q \log \lambda_{jlm}\bigr\} = \exp\bigl\{\Psi(\gamma_{jlm}) - \log \delta_{jlm}\bigr\}, \\
    \hat{\lambda}_{jlm} &= \mathbb{E}_q \lambda_{jlm} = \gamma_{jlm} / \delta_{jlm}. 
    \end{align*}
    \begin{align}
\text{Therefore, } b_{tjlm}^* &= \begin{cases} \exp\bigl\{\Psi(\zeta_{jl0}) - \Psi(\zeta_{jl\cdot})\bigr\} &\mbox{ if } m=0, \\
    %\\
    \exp\bigl\{ \Psi(\zeta_{jlm}) - \Psi(\zeta_{jl\cdot}) + \Psi( \gamma_{jlm}) - \log\delta_{jlm} - y_{tl} \frac{\gamma_{jlm}}{\delta_{jlm}} \bigr \} &\mbox{ if } m>0,
    \end{cases} \nonumber\\
    \text{and } b_{tj}^* &= \prod_{l=1}^L \sum_{m=0}^M b_{tjlm}^*. \nonumber
\end{align}
\endgroup
The quantities $a_{1j}^*$, $a_{jk}^*$ and $b_{tj}^*$ can be used as part of the Forward-Backward Algorithm to get our desired variational posterior estimates for the state probabilities as well as the cluster assignment probabilities. Implementation details of the variational Forward-Backward Algorithm is provided in Appendix \ref{s:VBFB}. The updates to the variational posterior on the latent variables are:
\begingroup
\allowdisplaybreaks
\begin{align*}
    q_{tj} &= \frac{\Tilde{F}_{tj}\cdot \Tilde{B}_{tj}}{\sum_{k=1}^K\Tilde{F}_{tk}\cdot\Tilde{B}_{tk}}, \\
q_{jk} &= \frac{\Tilde{F}_{tj}\cdot a_{jk}^*\cdot b_{t+1,k}^*\cdot \Tilde{B}_{t+1,k}}{\sum_{j=1}^K \sum_{k=1}^K \Tilde{F}_{tj}\cdot a_{jk}^*\cdot b_{t+1,k}^*\cdot \Tilde{B}_{t+1,k}}.
\end{align*}
\endgroup
where $\Tilde{F}_{tj}$ and $\Tilde{B}_{tj}$ are the scaled Forward and Backward variables respectively. The posterior update of $q_{1j}$ is the first entry of $q_{tj}$. The posterior for the mixture assignments for the $l^{th}$ location is given by
\begin{align*}
    q_{tjlm} \propto \begin{cases}
    1 &\mbox{ if } m=0, y_{tl}=0,\\
    0 &\mbox{ if } m>0, y_{tl}=0 \mbox{ or } m=0, y_{tl}>0, \\
    c_{jlm}^*f(y_{tl}|\lambda_{jlm}^*,\hat{\lambda}_{jlm}) &\mbox{ if } m>0, y_{tl} >0,
    \end{cases}
\end{align*}
where $ c_{jlm}^*f(y_{tl}|\lambda_{jlm}^*,\hat{\lambda}_{jlm}) = \exp\bigl\{ \Psi(\zeta_{jlm}) - \Psi(\zeta_{jl\cdot}) + \Psi( \gamma_{jlm}) - \log\delta_{jlm} - y_{tl} \frac{\gamma_{jlm}}{\delta_{jlm}} \bigr \}$.

\paragraph{Stochastic variational Bayes:} For SVB, we partition the data into $N$ years of $D$ days each, and assume exchangeability of the emission distributions for each year. Iterations of the VBEM are carried out on a minibatch randomly sampled using the sampling scheme described in \citet[Chapter 3.8]{ReetamThesis}. The sample is of size $D$; in the context of our current application, $D=92$ and $N=20$. The VBE step remains unchanged, and the VBM step is a stochastic gradient ascent of step size $\tau$. The expectations of the latent variables in the VBM step are scaled up by a factor of $N$ to reflect the entire data. The hyperparameter updates in the VBM step for the $i^{th}$ iteration for step size $\tau_i$ is given by:
\begingroup
\allowdisplaybreaks
\begin{align*}
        \xi_j^{(i)} &\leftarrow \bigl(1-\tau_i \bigr)\xi_j^{(i-1)} + \tau_i \bigl( \xi_j^{(0)} + q_{1j}\bigr),\\
        \zeta_{jl0}^{(i)} &\leftarrow \bigl(1-\tau_i \bigr)\zeta_{jl0}^{(i-1)} + \tau_i \bigl( \zeta_{jl0}^{(0)} + N \cdot \sum_{t=1}^D q_{tj}q_{tjl0} \bigr), \\
        \zeta_{jlm}^{(i)} &\leftarrow \bigl(1-\tau_i \bigr)\zeta_{jlm}^{(i-1)} + \tau_i \bigl( \zeta_{jlm}^{(0)} + N \cdot \sum_{t=1}^D q_{tj}q_{tjlm} \bigr), \\
        \gamma_{jlm}^{(i)} &\leftarrow \bigl(1-\tau_i \bigr)\gamma_{jlm}^{(i-1)} + \tau_i \bigl( \gamma_{jlm}^{(0)} + N \cdot \sum_{t=1}^D q_{tj}q_{tjlm} \bigr), \\
        \delta_{jlm}^{(i)} &\leftarrow \bigl(1-\tau_i \bigr)\delta_{jlm}^{(i-1)} + \tau_i \bigl( \delta_{jlm}^{(0)} + N \cdot \sum_{t=1}^D q_{tj}q_{tjlm}y_{tl} \bigr), \\
        \alpha_{jk}^{(i)} &\leftarrow \bigl(1-\tau_i \bigr)\alpha_{jk}^{(i-1)} + \tau_i \bigl(\alpha_{jk}^{(0)} + N \cdot \sum_{t=1}^{D-1} q_{jk} \bigr) ,
    \end{align*}
    where $j,k=1,\ldots,K$, $l=1, \ldots, L$, and $m=1,\ldots,M$. 
    
To assess the convergence of the VBEM algorithm, we compute and track the ELBO at each iteration. The ELBO for our model can be expressed as:
\begin{align*}
    ELBO(q) &= \mathbb{E}_{q(s,r)}\log p(y,s,r) + \mathbb{E}_{q(\Theta)}\log p(\Theta) + H\bigl(q(s,r)\bigr) - \mathbb{E}_{q(\Theta)}\log q(\Theta),
\end{align*}
where $H\bigl(q(s,r)\bigr)$ is the entropy of the variational posterior distribution over the latent variables. This simplifies to \citep{Beal03variationalalgorithms,Ji2006VariationalBF}:
\begin{align}
    \begin{split}
    ELBO(q) &= \log q(y|\Theta) - KL\bigl(q(\pi_1)\parallel p(\pi_1)\bigr) - KL\bigl(q(A)\parallel p(A)\bigr) \\
    & \text{\hspace{35mm}}- KL\bigl(q(C) \parallel p(C)\bigr) - KL \bigl(q(\Lambda) \parallel p(\Lambda)\bigr),
\end{split}
\end{align}
where the first term on the right hand side is calculated as part of the forward algorithm in \eqref{eqn:logprob}.
This relationship is used to compute the ELBO at each iteration, and we declare convergence once the change in ELBO falls below a desired threshold. 

\section{Results}
\subsection{Simulation Study}
We generate 1800 time steps of data from the proposed HMM at $L=3$ locations, with the number of states assumed known and fixed at $K=3$. At every location, positive precipitation is generated from a mixture of two exponential distributions, i.e. $M=2$. Conditional on the state, precipitation is independently distributed at the 3 locations. The true parameter values are:
\begin{enumerate}
    \item Initial probability vector $\pi_1 = (0.38,0.34,0.28)$ and transition matrix
\begin{align*}
A = \begin{bmatrix}
        0.60 & 0.30 & 0.10 \\
        0.20 & 0.50 & 0.30 \\
        0.30 & 0.20 & 0.50
\end{bmatrix}
\end{align*}
\item Matrices $C_1, C_2,$ and $C_3$ of mixture probabilities for the 3 locations
\begin{align*}
&C_1 = \begin{bmatrix}
        0.10 & 0.60 & 0.30 \\
        0.20 & 0.40 & 0.40 \\
        0.30 & 0.40 & 0.30
\end{bmatrix}, & C_2 &= \begin{bmatrix}
    0.20 & 0.70 & 0.10 \\
    0.40 & 0.20 & 0.40 \\
    0.50 & 0.20 & 0.30
    \end{bmatrix}, & C_3 &= \begin{bmatrix}
      0.20 & 0.60 & 0.20 \\
        0.50 & 0.30 & 0.20 \\
        0.60 & 0.20 & 0.20
\end{bmatrix}
\end{align*}
\item Matrices $\Lambda_1, \Lambda_2$, and $\Lambda_3$ with exponential distribution rate parameters for the 3 locations
\begin{align*}
&\Lambda_1 = \begin{bmatrix}
    0.08 & 1 \\
    0.60 & 5 \\
    1.00 & 8
    \end{bmatrix}, & \Lambda_2 &= \begin{bmatrix}
    0.05 & 1 \\
    0.50 & 4 \\
    1.00 & 10
    \end{bmatrix}, & \Lambda_3 &= \begin{bmatrix}
    0.10 & 1 \\
    0.10 & 5 \\
    0.90 & 6
    \end{bmatrix}.
\end{align*}
\end{enumerate}
The values of the mixture component assignments and the exponential rates are ordered such that state 1 corresponds to the wettest rainfall regime, and state 3 corresponds to the driest rainfall regime. This ensures model identifiability. For $l=1,2,\mbox{ and }3$, the rows of $C_l$ and $\Lambda_l$ correspond to the parameter values for each state.

We keep our prior specifications as broad as possible, and assign symmetric Dirichlet priors to $\pi_1$ and \textit{A}. The concentration of a Dirichlet distribution is defined as the sum of its parameters, and it indicates the amount of weight we put on the prior. We fix the concentration of $p(\pi_1)$ at 1, and the concentration for each row of $p(A)$ at 10. For each location $l$, the rows of $C_l$ have Dirichlet priors, and elements of $\Lambda_l$ have Gamma priors. The parameters for each location are assigned identical priors. The priors for the rows of $C_l$ are parameterized by $\zeta^{(0)}_l$. Similarly, the Gamma priors of the elements of $\Lambda_l$ have shape $\gamma^{(0)}$ and rate $\delta^{(0)}$. They are assigned the following values:
\begin{align*}
\zeta^{(0)}_l &= \begin{bmatrix}
3.0 & 4.0 & 3.0 \\
3.0 & 3.5 & 3.5 \\
4.0 & 3.0 & 3.0
\end{bmatrix}  &\gamma^{(0)}_l &= \begin{bmatrix}
    0.5 & 2 \\
    1.5 & 9 \\
    2.0 & 16
    \end{bmatrix}  &\delta^{(0)}_l &= \begin{bmatrix}
    2 & 2 \\
    2 & 2 \\
    2 & 2
    \end{bmatrix}
\end{align*}
These assignments follow the reasoning that wetter states will have lower exponential rates and higher mixture probabilities for exponential components, while drier states will have higher rates and more weight placed on the dry component corresponding to $m=0$.

The VBEM algorithm converges at 752 iterations when we set the threshold to $10^{-9}$. The posterior for the initial state probability is $\Tilde{\pi}_1 = (0.21, 0.62, 0.17)$. The posterior for the transition probability matrix is
\begin{align*}
&\Tilde{A} = \begin{bmatrix}
        0.62 & 0.23 & 0.15 \\
        0.24 & 0.44 & 0.32 \\
        0.29  & 0.34  & 0.37
\end{bmatrix},
\end{align*}
and the posterior distributions of the mixture components and exponential rates are
\begin{align*}
&\Tilde{C}_1 = \begin{bmatrix}
        0.10 & 0.61 & 0.29 \\
        0.24 & 0.47 & 0.29 \\
        0.29 & 0.43 & 0.29
\end{bmatrix} & \Tilde{C}_2 &= \begin{bmatrix}
    0.19 & 0.70 & 0.11 \\
    0.36 & 0.23 & 0.41 \\
    0.55 & 0.21 & 0.24
    \end{bmatrix} & \Tilde{C}_3 &= \begin{bmatrix}
      0.20 & 0.65 & 0.15 \\
        0.58 & 0.32 & 0.10 \\
        0.49 & 0.21 & 0.30
\end{bmatrix},\\
&\Tilde{\Lambda}_1 = \begin{bmatrix}
    0.09 & 1.06 \\
    0.71 & 8.06 \\
    1.25 & 7.64
    \end{bmatrix} & \Tilde{\Lambda}_2 &= \begin{bmatrix}
    0.05 & 1.27 \\
    0.51 & 5.13 \\
    1.00 & 9.78
    \end{bmatrix} & \Tilde{\Lambda}_3 &= \begin{bmatrix}
    0.10 & 1.62 \\
    0.12 & 4.80 \\
    1.00 & 7.42
    \end{bmatrix}.
\end{align*}
The posterior distributions are generally close to the true values, with the exception of the initial distribution. Moreover, the posteriors for the wet states and for heavy rainfall are better estimated in this study. The quality of the estimates do not seem to be noticeably affected by the broad priors, and as long as the model is identifiable, we do not anticipate this being a problem.

\begin{table}
\centering
\caption{Comparison of true HMM states and states decoded using the Viterbi algorithm after parameter estimation.}
\label{tab:viterbisim}
\begin{tabular}{@{}rllll@{}}
\toprule
\multicolumn{1}{l}{} &  & \multicolumn{3}{c}{True states} \\ 
\multicolumn{1}{l}{} & \multicolumn{1}{l}{} & 1 & 2 & 3 \\ \midrule
\multirow{3}{*}{\begin{tabular}[c]{@{}r@{}}Decoded\\ states\end{tabular}} & \multicolumn{1}{l|}{1} & 631 & 71 & 11 \\
 & \multicolumn{1}{l|}{2} & 50 & 363 & 207 \\
 & \multicolumn{1}{l|}{3} & 24 & 162 & 281 \\ \bottomrule
\end{tabular}
\end{table}
One of the goals for our study is to characterize the underlying weather regimes present in the data, which requires the identification, or decoding, of the hidden states. We apply the Viterbi algorithm \citep{Viterbi} to decode the most likely sequence of states that could have generated the data based on our fitted model. Table \ref{tab:viterbisim} contains a comparison of the true and decoded states for the study. We note that State 1 has the highest classification accuracy of 89.5\%, followed by State 2 with an accuracy of 60.9\% and State 3 with 56.3\%. This is consistent with our observations about the posterior estimates, that the wettest state (State 1) has the most accurate posterior mean. Overall, we see that State 1 is well discriminated from the others, whereas States 2 and 3 have much more mis-classification error. This is not unexpected if our data arises from a wet season, as this simulation study is designed to reflect. It is likely that a study of the dry seasons could lead to the highest classification accuracy for the dry state. 

\begin{figure}
    \centering
    \begin{subfigure}[b]{0.45\linewidth}
     \includegraphics[width=\linewidth]{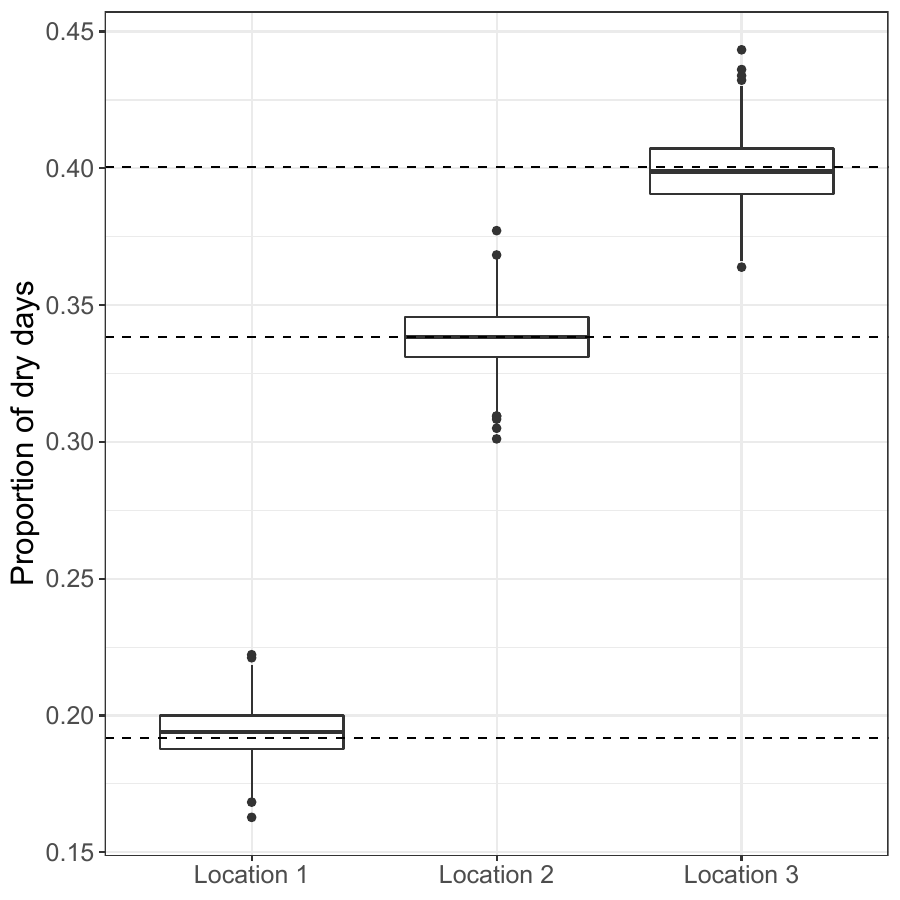}
\caption{Distribution of the proportion of dry days at each location.}
    \label{fig:sim_drydays}
    \end{subfigure}
    \hfill
    \begin{subfigure}[b]{0.45\linewidth}
        \ \includegraphics[width=\linewidth]{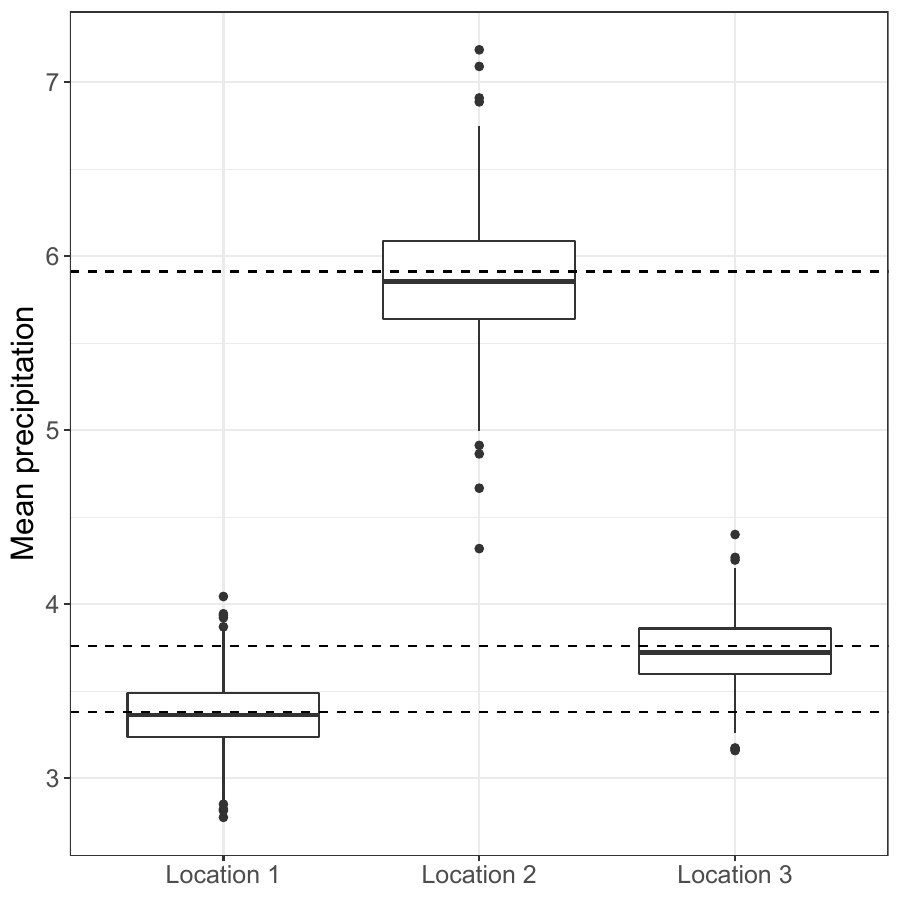}
\caption{Distribution of mean precipitation (mm) at each location.}
    \label{fig:sim_wetdays}
    \end{subfigure}
        \caption{Proportion of dry days and mean precipitation (mm) estimated from 1000 datasets each generated using the variational posterior estimates. Each dataset consists of a sequence of 1800 days. The dotted lines represent the values in the training data.}
    \label{fig:sim_boxplot}
\end{figure}

To verify whether the model can estimate key sample statistics of interest, we generated 1000 datasets each 1800 days long from the variational posterior. We compute the proportion of dry days and mean rainfall for wet days from each of these 1000 datasets, and compare them with the values in the training data. Figure \ref{fig:sim_boxplot} plots the distribution of the proportion of dry days and mean precipitation based on data generated from the posterior at each of the 3 locations. The dotted lines in the figures represent the values in the original data. In all cases, the posterior estimates are centered around the values in the training (original) data. In Figure \ref{fig:sim_drydays}, the root mean square error (RMSE) between the posterior estimates and the training data values is 0.01 at all 3 locations. Similarly, the RMSE in $mm$ between the mean precipitation estimates from the posterior and the training data values is $0.2, 0.35, \,\mbox{and } 0.19$ at the three locations. The fitted model functions well as an SPG, and can replicate the marginal distributions of precipitation at least in cases where the number of locations is not too large.

\subsection{HMM for daily precipitation over the Chesapeake Bay watershed}

We now fit an HMM to daily precipitation data from GPM-IMERG over the Chesapeake Bay watershed using 20 years of wet season data. Our objectives are twofold - identify the underlying states and the weather regimes they correspond to, and construct an SPG to replicate the marginal and spatial characteristics of the remote sensing data. We fit a 3-state HMM to the data corresponding to the model presented in Section \ref{s:precipHMM}. Our model priors are nearly identical to what was used in the simulation study.
We assign symmetric Dirichlet priors to $\pi_1$ and \textit{A}. The prior $p(\pi_1)$ has a concentration of 1, and each row of the prior $p(A)$ has a concentration of 10. Without loss of generality, we order the states to correspond to heavy, medium, and low rainfall respectively. For each location $l = 1, \ldots, 1927$, precipitation is specified as a mixture with a point mass at zero and two exponential distributions for positive precipitation. The prior parameters $\zeta^{(0)}_l$, $\gamma^{(0)}_l$, and $\delta^{(0)}_l$ are defined as before and assigned the following values:
\begin{align*}
\zeta^{(0)}_l &= \begin{bmatrix}
3.0 & 4.0 & 3.0 \\
3.0 & 3.5 & 3.5 \\
4.0 & 3.0 & 3.0
\end{bmatrix},  &\gamma^{(0)}_l &= \begin{bmatrix}
    0.5 & 2 \\
    1.5 & 5 \\
    2.0 & 10
    \end{bmatrix}, &\delta^{(0)}_l &= \begin{bmatrix}
    2 & 2 \\
    2 & 2 \\
    2 & 2
    \end{bmatrix}.
\end{align*}
Model identifiability is ensured by ordering the components such that wetter states will have lower exponential rates and higher mixture probabilities for the exponential components, while drier states will have higher rates and more weight placed on the dry component corresponding to $m=0$. 

To fit the model, we ran SVI optimization with step sizes $\tau_i = (1+i)^{-0.9}$ for 500 iterations. This was followed by 50 iterations of CAVI using the entire data to ensure convergence of the algorithm. The fitted model has a posterior initial probability $\tilde{\pi}_1 = c(0.11,0.44,0.45)$ and the transition probability matrix
\begin{align*}
    \tilde{A} = \begin{bmatrix}
    0.40 & 0.38 & 0.22 \\
    0.32 & 0.38 & 0.30 \\
    0.12 & 0.32 & 0.56
    \end{bmatrix}.
\end{align*}
We note that the two lowest probabilities in the transition matrix occur when the driest state transitions to the wettest state (0.12), and vice versa (0.22). States 1 and 3 tend to transition between each other through state 2 most of the time. 

\begin{figure}
    \centering
     \includegraphics[width=\linewidth]{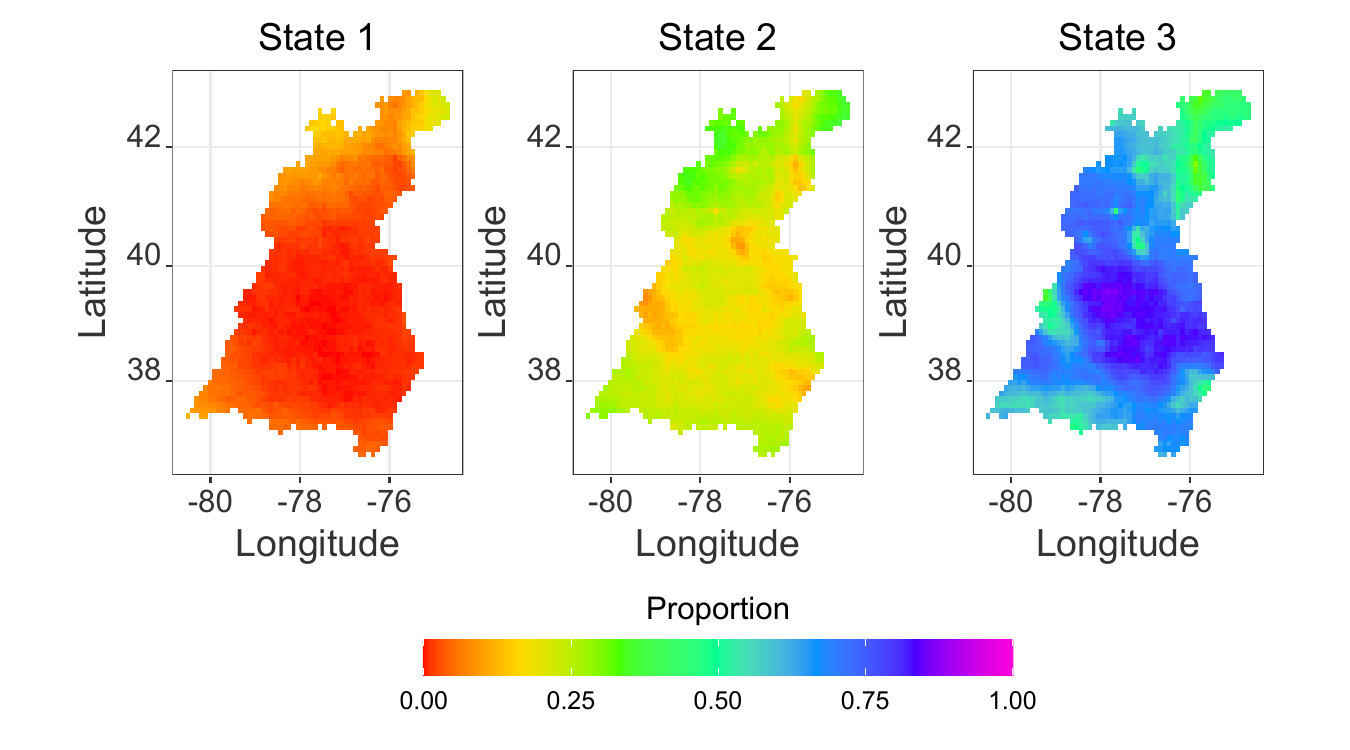}
\caption{\small Distribution of the proportion of dry days for each of the 3 HMM states for the Chesapeake Bay watershed data.}
    \label{fig:cbay_obs_dry_state}
    \end{figure}
    \begin{figure}
        \ \includegraphics[width=\linewidth]{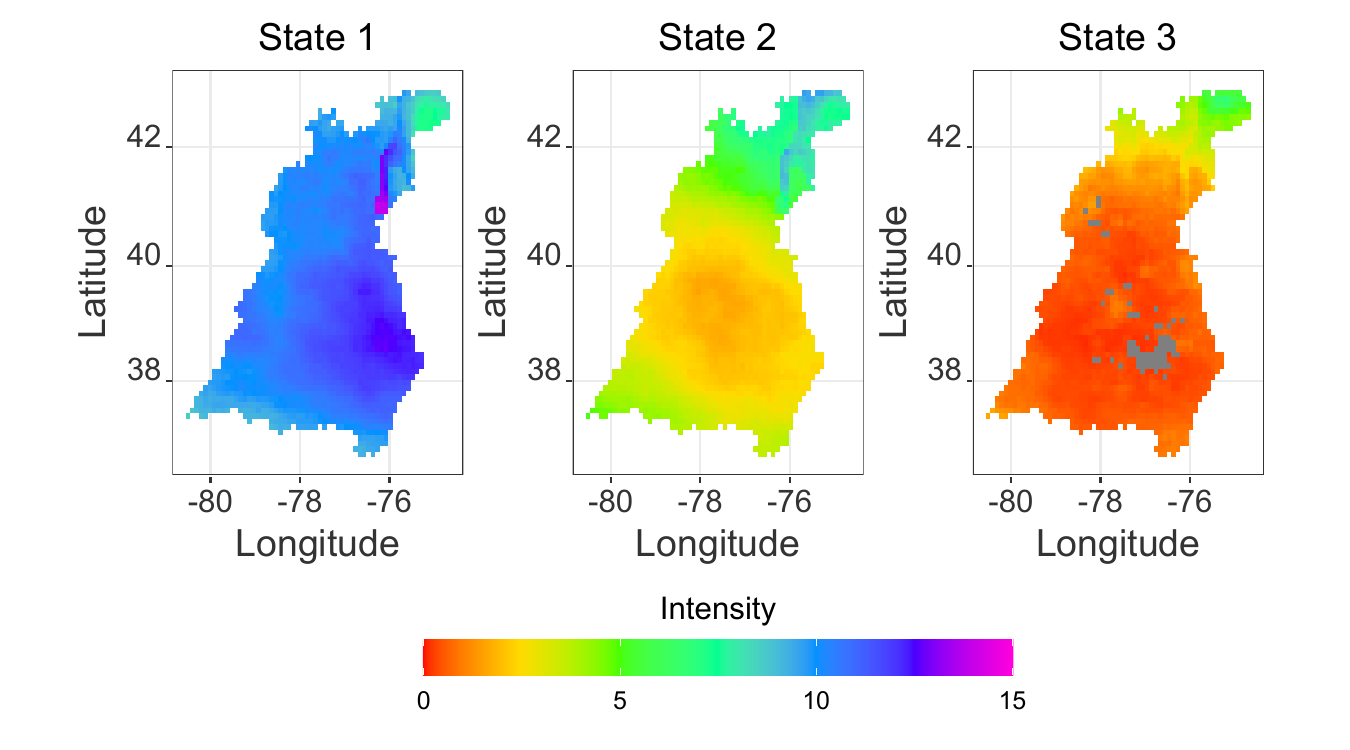}
\caption{Distribution of mean precipitation intensity (mm) for each of the 3 HMM states for the Chesapeake Bay watershed data. Gray pixels denote areas with no data.}
    \label{fig:cbay_obs_int_state}
\end{figure}

\begin{table}
\centering
\caption{Percentage of days in each month that correspond to the decoded Viterbi states for the HMM.}
\label{tab:viterbicbay}
\begin{tabular}{@{}rllll@{}}
\toprule
\multicolumn{1}{l}{} &  & \multicolumn{3}{c}{Month} \\ 
\multicolumn{1}{l}{} & \multicolumn{1}{c}{} & Jul & Aug & Sep \\ \midrule
\multirow{3}{*}{\begin{tabular}[c]{@{}r@{}}Decoded\\ states\end{tabular}} & \multicolumn{1}{l|}{1} & 27 & 25 & 27 \\
 & \multicolumn{1}{l|}{2} & 34 & 39 & 34 \\
 & \multicolumn{1}{l|}{3} & 38 & 36 & 38 \\ \bottomrule
\end{tabular}
\end{table}

Running the Viterbi decoding on the fitted model, we find that 492 days are estimated to be in State 1, 659 in State 2, and 689 in State 3. As we have seen in Figure \ref{fig:imerg}, the heaviest precipitation occurs in the northern part of the watershed, and most of the remaining area gets significantly less rainfall. Table \ref{tab:viterbicbay} breaks down the distribution of states by month. Each cell represents the percentage of days in a particular month that the HMM was in that state, averaged over the 20 years of data. Each column thus adds up to 100\%. We see that July and September have similar state distributions. August on the other hand has the highest proportion of days in State 2. 
%\clearpage

Figures \ref{fig:cbay_obs_dry_state} and \ref{fig:cbay_obs_int_state} plot the spatial distributions of the proportion of dry days and the mean precipitation intensity over the watershed for each of the 3 states respectively. We define the mean intensity as the mean precipitation on days when it has rained. State 1 corresponds to high precipitation amounts across the watershed except for a small section in the north. State 3 is the driest state, and is the only state where some locations have had  zero precipitation days. This is denoted by the greyed out pixels in Figure \ref{fig:cbay_obs_int_state} where there is no non-zero precipitation data available based on the fitted model. This is coupled with the highest proportion of dry days for the same area seen in Figure \ref{fig:cbay_obs_dry_state}. Interestingly, the northern part of the watershed which gets some of the lowest precipitation for State 1, gets the highest precipitation in State 3. This suggests that the northernmost part of the watershed could have different underlying weather patterns compared to the rest of the Bay. Finally, State 2 has precipitation patterns that are somewhere between States 1 and 3. In Figure \ref{fig:cbay_obs_dry_state}, we see areas of the south-west and the east that are wetter than the rest of the watershed in States 2 and 3, and in Figure \ref{fig:cbay_obs_int_state}, State 2 outlines the driest part of the watershed which is the same area with zero precipitation days under State 3.
\begin{figure}
    \centering
    \begin{subfigure}[b]{0.47\linewidth}
     \includegraphics[width=\linewidth]{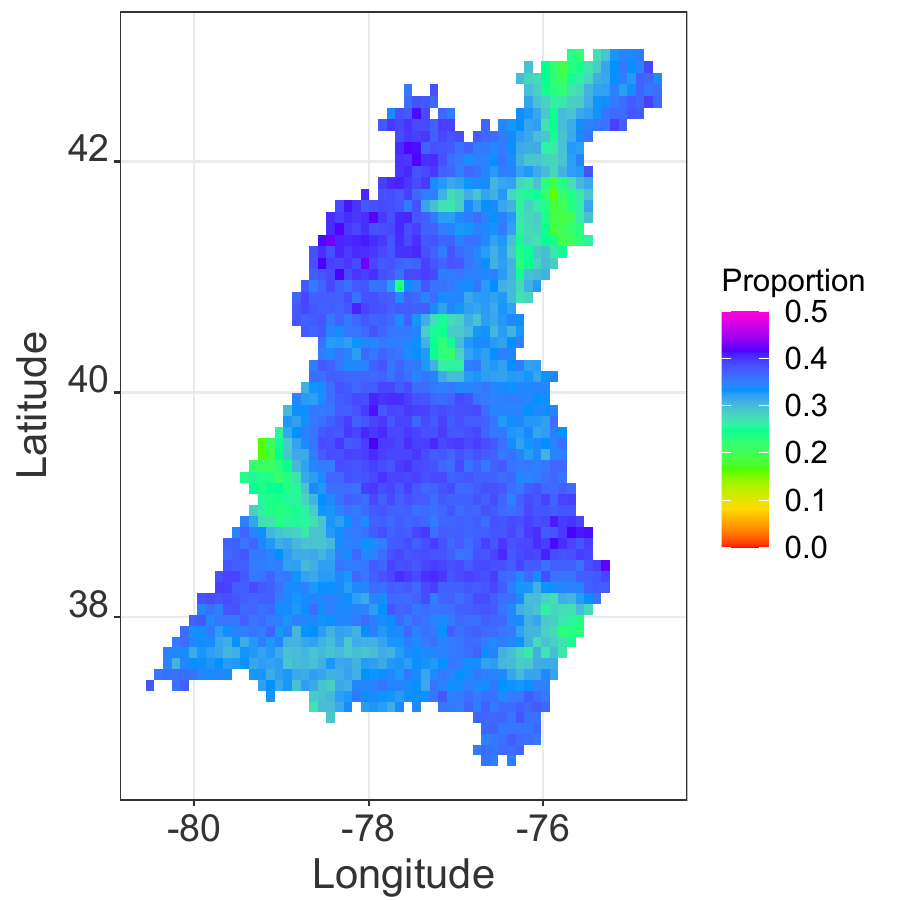}
\caption{Proportion of dry days at each location.}
    \label{fig:cbay_sim_dry}
    \end{subfigure}
    \hfill
    \begin{subfigure}[b]{0.47\linewidth}
        \includegraphics[width=\linewidth]{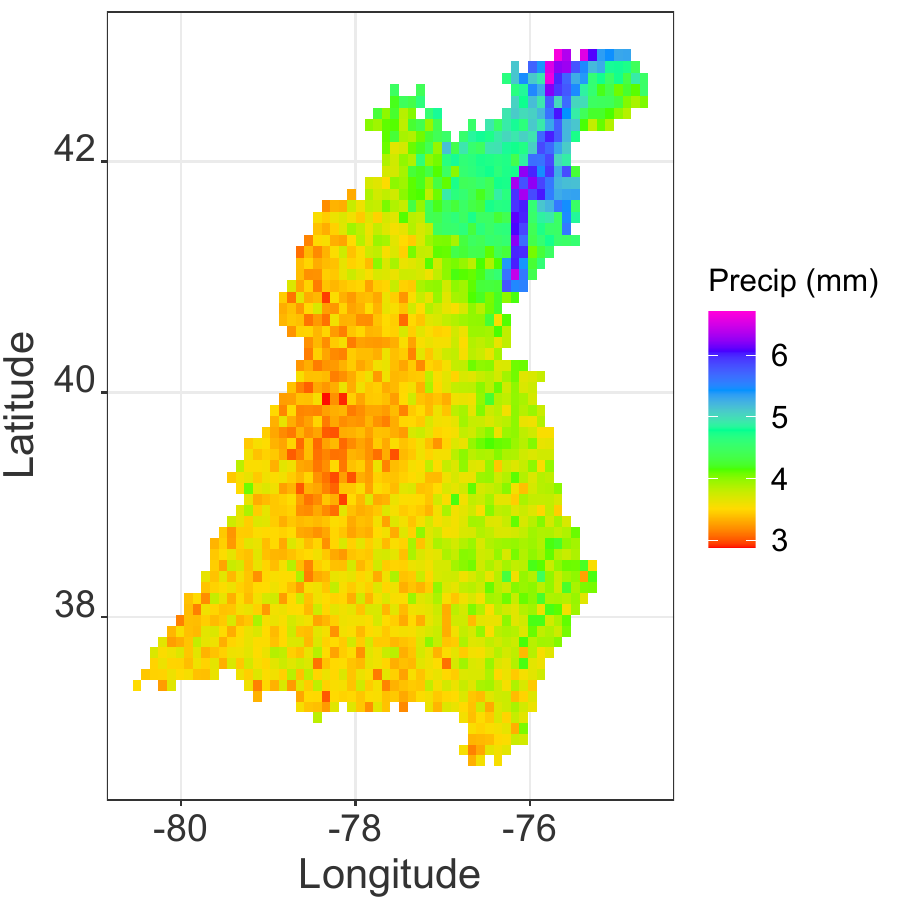}
\caption{Mean precipitation (mm) at each location.}
    \label{fig:cbay_sim_wet}
    \end{subfigure}
        \caption{Proportion of dry days and mean daily precipitation during Jul--Sep for the Chesapeake Bay watershed based on synthetic data generated from a fitted 3-state HMM.}
    \label{fig:cbay_sim_1}
\end{figure}

To test this model as a precipitation generator, we generated 1840 days of synthetic data from the 3-state HMM. The generated data replicates the spatial and temporal structure of the GPM-IMERG data. Figure \ref{fig:cbay_sim_1} shows a plot of total precipitation at each grid point over the 3 months of Jul--Sep averaged over 20 years, based on the synthetic data simulated from the model. While the plot is noisier compared to the corresponding plot of the historical data in Figure \ref{fig:imerg}, it is largely able to recreate seasonal precipitation patterns at individual locations. 

\begin{figure}
    \centering
    \begin{subfigure}[b]{0.45\linewidth}
     \includegraphics[width=\linewidth]{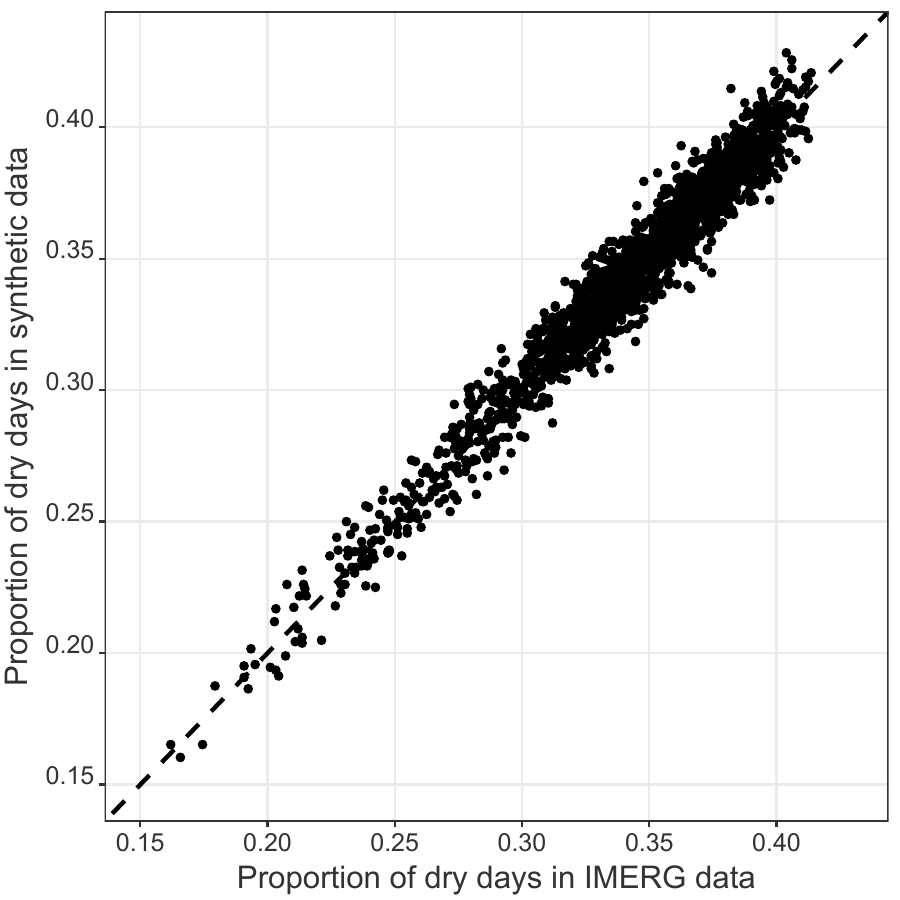}
\caption{Proportion of dry days in historical and synthetic data for each location.}
    \label{fig:dry_rmse}
    \end{subfigure}
    \hfill
    \begin{subfigure}[b]{0.45\linewidth}
        \ \includegraphics[width=\linewidth]{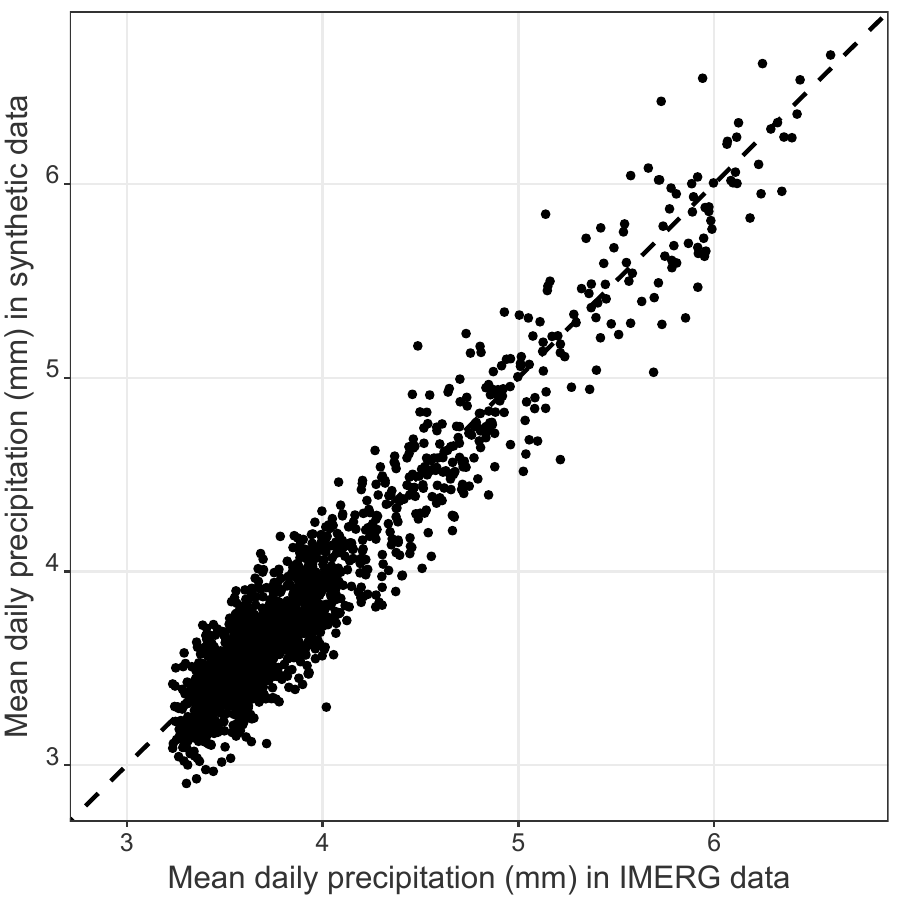}
\caption{Mean daily precipitation (mm) in historical and synthetic data for each location.}
    \label{fig:wet_rmse}
    \end{subfigure}
        \caption{Proportion of dry days and mean daily precipitation (mm) during Jul--Sep for the Chesapeake Bay watershed based on historical GPM-IMERG data and synthetic data generated from the fitted 3-state HMM.}
    \label{fig:cbay_sim_2}
\end{figure}

Figure \ref{fig:dry_rmse} plots the proportion of dry days averaged over 20 years at each location based on historical IMERG data on the $x$-axis and synthetic data from the HMM on the $y$-axis, which has an RMSE of 0.009. Similarly, Figure \ref{fig:wet_rmse} plots the mean daily precipitation averaged over 20 years at each location based on historical IMERG data on the $x$-axis and synthetic data from the HMM on the $y$-axis, with an RMSE of 0.181 mm. In both cases, the line through the middle of the plot corresponds to $y=x$. In both plots, the points show a linear pattern, and aside from a slight underestimation of low values in Figure \ref{fig:wet_rmse}, the 3-state HMM with VB parameter estimation is able to reproduce seasonal characteristics of precipitation at each location.

%%%%%%%%%%%%%%%%%%%%%%%%%%%%%%%%%%%%%%%%%%
\section{Discussion}

The spatiotemporal analysis of remote sensing data requires approaches which can account for spatial surfaces which exhibit significant heterogeneity, while retaining computational efficiency. This is complicated further by datasets like precipitation whose distributions are not only non-normal, but are semi-continuous in nature. In this paper, we have developed a workflow for efficient parameter estimation in HMMs for semi continuous emissions using VB. The Bayesian approach allows us to incorporate prior information, and the variational approximation enables fast computations. We are able to exploit the seasonal structure of the data to implement stochastic optimization which reduces computation time further by making it a function of the number of locations and season length, instead of the entire data length. 

We use the variational HMM to study precipitation dynamics for the wet season months of July--September over the Chesapeake Bay watershed. We note distinct precipitation patterns associated with the 3 states, and that different parts of the watershed are often affected differently under the 3 states even if their overall seasonal precipitation is similar. To determine the model's capability as an SPG, we generated a synthetic dataset from the model and compared two metrics at each location - the proportion of dry days, and the mean daily precipitation. A linear trend is seen between the historical and synthetic data for both metrics as well as low RMSE values. This indicates a well specified model which can capture and replicate the mean behavior of the data.

One of the assumptions made in our work is stationarity over time. There are two types of stationarity to consider here. The first is stationarity within a year's data. This is addressed by considering only seasonal data - an approach previous studies have also taken. However, there is the additional assumption that the states are unchanged from year-to-year, and the emission distributions for different years are interchangeable. This assumption has been exploited for stochastic optimization. The model can be relaxed to accommodate underlying changes in climate if a longer data record is available. In particular, any prediction or forecasting needs to consider climate model outputs to reliably track the change in precipitation patterns over time.

Future work will focus on explicitly parameterizing the spatial dependence by augmenting the emission distribution with a copula. While the current model accommodates the spatial heterogeneity in the data implicitly through the Markov chain, explicitly parameterizing the dependence will help identify states or areas where it plays the biggest role. This is especially important for precipitation, whose spatial distribution is not as smooth as, say, temperature. We have used copulas in previous work to estimate spatial correlations in an HMM \citep{MajumderJSM2020}, but it has been in the context of maximum-likelihood estimation using the Baum-Welch algorithm and not in a Bayesian setup. We would also like to incorporate downscaled climate model outputs as covariates for the model as in \citet{RobertsonKirshnerSmyth2004}. This has two benefits - it will allow us to tune its behavior, and we would also be able to specify a more sophisticated non-homogeneous HMM (NHMM) where the Markov chain parameters can be made to vary by month, or even by season. An NHMM, however, has additional computational complexity; the transition matrix parameters tend to be linked to covariates by a probit or logistic link function, and the resulting model is no longer in the conjugate exponential family. Finally, we would like to explore model selection in more detail, both in terms of the model size as well as our choice of distribution used to specify positive precipitation.

%%%%%%%%%%%%%%%%%%%%%%%%%%%%%%%%%%%%%%%%%%
\vspace{6pt} 

\section*{Acknowledgements}
The hardware used in the computational studies is part of the UMBC High Performance Computing Facility (HPCF). The facility is supported by the U.S. National Science Foundation through the MRI program (grant nos.~CNS--0821258, CNS--1228778, and OAC--1726023)
and the SCREMS program (grant no.~DMS--0821311), with additional substantial support from the University of Maryland, Baltimore County (UMBC). 
See \url{hpcf.umbc.edu}
for more information on HPCF and the projects using its resources.
% Additional acknowledgments for co-authors:
Co-author Reetam Majumder was supported by the Joint Center for Earth Systems Technology and by the HPCF as a Research Assistant.

\begin{singlespace}
	\bibliographystyle{rss}
	\bibliography{rain.bib}
\end{singlespace}
\begin{appendices}
%\clearpage
\section{Variational Forward-Backward Algorithm}\label{s:VBFB}
\begin{comment}
 Comparing this expression with the canonical form for the conjugate exponential family, we arrive at the following expressions for the natural parameters $\phi(\Theta)$, their sufficient statistics $u(s,y,r)$, and the hyperparameters $\nu^{(0)}$:
\begin{align}\label{eqn:parameters2}
    \phi(\Theta) &= \begin{bmatrix}
                    \log\pi_{1j} \\
                    \log c_{jl0} \\
                    \log c_{jlm} \\
                    \log \lambda_{jlm} \\
                    \lambda_{jlm} \\
                    \log a_{jk}
           \end{bmatrix}, & u(s,y,r) &= \begin{bmatrix}
                                        s_{1j} \\
                                        s_{tj}r_{tjl0} \\
                                        s_{tj}r_{tjlm} \\
                                        s_{tj}r_{tjlm} \\
                                        y_{tl} s_{tj}r_{tjlm} \\
                                        s_{tj}s_{t+1,k}
                                        \end{bmatrix}, & \nu^{(0)} &= \begin{bmatrix}
                                \xi_j^{(0)} - 1 \\
                                \zeta_{jl0}^{(0)}-1 \\
                                \zeta_{jlm}^{(0)}-1 \\
                                \gamma_{jlm}^{(0)}-1 \\
                                \delta_{jlm}^{(0)} \\
                                \alpha_{jk}^{(0)}-1
                                   \end{bmatrix},
\end{align}
\endgroup
for $m = 1, \ldots, M, j = 1, \ldots, K, k = 1, \ldots, K$. 
\end{comment}

The Forward Variable is defined as the joint probability of the partial observation sequence up to a time \textit{t}, and the state $s_t$ at that time point:
\begin{align*}
    F_{tj} &= p(y_1,\ldots, y_t, s_t=j).
\end{align*}
\begin{enumerate}
    \item \textbf{Initialization}: For all $j=1, \ldots, K$, define
    \begin{align*}
        F_{1j} &= \pi_1\cdot  p(y_1|s_1=j), \\
        c_1 &= \frac{1}{\sum_{j=1}^K F_{1j}} \text{\hspace{1mm} and normalize}\\
        \Tilde{F}_{1j} &= c_1 \cdot F_{1j}.
    \end{align*}
    \item \textbf{Recursion}: for $t=2, \ldots, T$ and for each state $k = 1, \ldots, K$, use the recursion
    \begin{align*}
        F_{tk} &= \biggl[ \sum_{j=1}^K \Tilde{F}_{t-1,j}\cdot p(s_t=k|s_{t-1} = j)\biggr]p(y_t|s_t=k) \text{\hspace{1mm} and normalize}\\
        \Tilde{F}_{tj} &= c_t \cdot F_{tk} \text{\hspace{1mm} where}\\
        c_t &= \frac{1}{\sum_{j=1}^K F_{tj}}.
    \end{align*}
\end{enumerate}
Note that $\Tilde{F}_{tj} = (\prod_{\tau=1}^t c_\tau) F_{tj}$. Using the definitions provided, this gives us
\begin{align}\label{eqn:logprob}
    q(y|\Theta) = \sum_{j=1}^K F_{Tj} = \frac{1}{\prod_{t=1}^Tc_t},
\end{align}
where $q(y|\Theta)$ is the normalizing constant for the variational posterior of the latent variables. Recall that the Forward Algorithm is used as part of the E-step of the optimization process, with the values of the parameters in $\Theta$ set to their means, i.e., $\Theta\equiv \tilde{\Theta}$. Thus $q(y|\Theta)$ can equivalently also be expressed as $p(y|\tilde{\Theta})$.

The Backward Variable is defined as the probability of generating the last $T-t$ observations given that the system is in state \textit{j} at time \textit{t}:
\begin{align*}
    B_{tj} = p(y_{t+1}, \ldots, y_{T} | s_t=j).
\end{align*}
The Backward Algorithm has similar steps but works its way back from the final time point.
\begin{enumerate}
    \item \textbf{Initialization}: For each state \textit{j}, set
    \begin{align*}
        B_{Tj} &= 1 \text{\hspace{1mm}, and} \\
        \Tilde{B}_{Tj} &= c_T \cdot B_{Tj}.
    \end{align*}
    \item \textbf{Recursion}: for $t=T-1, \ldots, 1$ and each state \textit{j}, calculate
    \begin{align*}
        B_{tj} &= \sum_{k=1}^K p(s_{t+1}=k|s_t=j)\cdot \Tilde{B}_{t+1,k}\cdot p(y_{t+1}|s_{t+1}=k), \\
        \Tilde{B}_{tj} &= c_t \cdot B_{tj}.
    \end{align*}
\end{enumerate}
The two algorithms can be run in parallel. Once both variables are calculated, we get
\begin{align*}
        q_s(s_t=j|y_1, \ldots, y_T) &\propto \Tilde{F}_{tj}\cdot \Tilde{B}_{tj},\text{ and } \\
q_s(s_t=j,s_{t+1}=k) &\propto \Tilde{F}_{tj}\cdot p(s_{t+1}=k|s_t=j)\cdot p(y_{t+1}|s_{t+1}=k)\cdot \Tilde{B}_{t+1,k}.
\end{align*}
\end{appendices}
\end{document}

%% file: commands.tex
\newcommand{\beq}{ \begin{equation}}
\newcommand{\eeq}{ \end{equation}}
\newcommand{\beqn}{ \begin{eqnarray}}
\newcommand{\eeqn}{ \end{eqnarray}}

\usepackage{caption}

%% file: rain.bbl
\begin{thebibliography}{35}
\expandafter\ifx\csname natexlab\endcsname\relax\def\natexlab#1{#1}\fi
\expandafter\ifx\csname url\endcsname\relax
  \def\url#1{\texttt{#1}}\fi
\expandafter\ifx\csname urlprefix\endcsname\relax\def\urlprefix{URL}\fi

\bibitem[{Attias(1999)}]{attias1999}
Attias, H. (1999) Inferring parameters and structure of latent variable models
  by variational {B}ayes.
\newblock In \textit{Proceedings of the Fifteenth Conference on Uncertainty in
  Artificial Intelligence}, UAI’99, 21–30. Morgan Kaufmann Publishers Inc.

\bibitem[{Baum(1972)}]{Baum1972}
Baum, L.~E. (1972) An inequality and associated maximization technique in
  statistical estimation for probabilistic functions of {M}arkov processes.
\newblock In \textit{Inequalities {III}: {P}roceedings of the Third Symposium
  on Inequalities}, 1--8. University of California, Los Angeles.

\bibitem[{Baum and Eagon(1967)}]{baum1967}
Baum, L.~E. and Eagon, J.~A. (1967) An inequality with applications to
  statistical estimation for probabilistic functions of {Markov} processes and
  to a model for ecology.
\newblock \textit{Bull. Amer. Math. Soc.}, \textbf{73}, 360--363.

\bibitem[{Baum and Petrie(1966)}]{baum1966}
Baum, L.~E. and Petrie, T. (1966) Statistical inference for probabilistic
  functions of finite state {M}arkov chains.
\newblock \textit{Ann. Math. Statist.}, \textbf{37}, 1554--1563.

\bibitem[{Baum et~al.(1970)Baum, Petrie, Soules and Weiss}]{baum1970}
Baum, L.~E., Petrie, T., Soules, G. and Weiss, N. (1970) A maximization
  technique occurring in the statistical analysis of probabilistic functions of
  {Markov} chains.
\newblock \textit{Ann. Math. Statist.}, \textbf{41}, 164--171.

\bibitem[{Baum and Sell(1968)}]{baum1968}
Baum, L.~E. and Sell, G.~R. (1968) Growth transformations for functions on
  manifolds.
\newblock \textit{Pacific J. Math.}, \textbf{27}, 211--227.

\bibitem[{Beal(2003)}]{Beal03variationalalgorithms}
Beal, M.~J. (2003) Variational algorithms for approximate {B}ayesian inference.
\newblock Ph.D. Thesis, Gatsby Computational Neuroscience Unit, University
  College London.

\bibitem[{Bellone et~al.(2000)Bellone, Hughes and Guttorp}]{bellone2000}
Bellone, E., Hughes, J. and Guttorp, P. (2000) A hidden {M}arkov model for
  downscaling synoptic atmospheric patterns to precipitation amounts.
\newblock \textit{Clim. Res.}, \textbf{15}, 1--12.

\bibitem[{Blei et~al.(2017)Blei, Kucukelbir and McAuliffe}]{bleietal2017}
Blei, D.~M., Kucukelbir, A. and McAuliffe, J.~D. (2017) Variational inference:
  A review for statisticians.
\newblock \textit{J. Am.\ Stat.\ Assoc.}, \textbf{112}, 859--877.

\bibitem[{Breinl et~al.(2017)Breinl, Di~Baldassarre, Girons~Lopez, Hagenlocher,
  Vico and Rutgersson}]{brienletal2017}
Breinl, K., Di~Baldassarre, G., Girons~Lopez, M., Hagenlocher, M., Vico, G. and
  Rutgersson, A. (2017) Can weather generation capture precipitation patterns
  across different climates, spatial scales and under data scarcity?
\newblock \textit{Sci.\ Rep.-UK}, \textbf{7}.

\bibitem[{Capp\'{e} et~al.(2005)Capp\'{e}, Moulines and Ryden}]{Cappeetal2005}
Capp\'{e}, O., Moulines, E. and Ryden, T. (2005) \textit{Inference in Hidden
  Markov Models (Springer Series in Statistics)}.
\newblock Berlin, Heidelberg: Springer-Verlag.

\bibitem[{Dempster et~al.(1977)Dempster, Laird and Rubin}]{DempsterElAl1977}
Dempster, A.~P., Laird, N.~M. and Rubin, D.~B. (1977) Maximum likelihood from
  incomplete data via the {EM} algorithm.
\newblock \textit{J.\ R.\ Stat. Soc. B}, \textbf{39}, 1--22.

\bibitem[{Foti et~al.(2014)Foti, Xu, Laird and Fox}]{NIPS2014_Fotietal}
Foti, N., Xu, J., Laird, D. and Fox, E. (2014) Stochastic variational inference
  for hidden {M}arkov models.
\newblock In \textit{Advances in Neural Information Processing Systems},
  vol.~27. Curran Associates, Inc.

\bibitem[{Ghahramani and Beal(2000)}]{Grahramanietal}
Ghahramani, Z. and Beal, M.~J. (2000) Propagation algorithms for variational
  {B}ayesian learning.
\newblock In \textit{13th International Conference on Neural Information
  Processing Systems}, NIPS’00, 486–492. MIT Press.

\bibitem[{Hoffman et~al.(2013)Hoffman, Blei, Wang and
  Paisley}]{JMLR:v14:hoffman13a}
Hoffman, M.~D., Blei, D.~M., Wang, C. and Paisley, J. (2013) Stochastic
  variational inference.
\newblock \textit{J.Mach.\ Learn.\ Res.}, \textbf{14}, 1303--1347.

\bibitem[{Holsclaw et~al.(2016)Holsclaw, Greene, Robertson and
  Smyth}]{Holsclaw2016}
Holsclaw, T., Greene, A.~M., Robertson, A.~W. and Smyth, P. (2016) A {B}ayesian
  hidden {M}arkov model of daily precipitation over {S}outh and {E}ast {A}sia.
\newblock \textit{J. Hydrometeorol.}, \textbf{17}, 3--25.

\bibitem[{Huffman et~al.(2019)Huffman, Stocker, Bolvin, Nelkin and Tan}]{IMERG}
Huffman, G.~J., Stocker, E.~F., Bolvin, D.~T., Nelkin, E.~J. and Tan, J. (2019)
  {GPM} {IMERG} final precipitation {L3} 1~day 0.1~degree $\times$ 0.1~degree
  {V06}.
\newblock Edited by Andrey Savtchenko, Greenbelt, MD, Goddard Earth Sciences
  Data and Information Services Center (GES DISC),
  \url{https://disc.gsfc.nasa.gov/datasets/GPM_3IMERGDF_06/summary}, accessed
  on Aug~28, 2020.

\bibitem[{Hughes and Guttorp(1994)}]{HughesGuttorp1994}
Hughes, J.~P. and Guttorp, P. (1994) Incorporating spatial dependence and
  atmospheric data in a model of precipitation.
\newblock \textit{J. Appl.\ Meteorol.}, \textbf{33}, 1503--1515.

\bibitem[{Ji et~al.(2006)Ji, Krishnapuram and Carin}]{Ji2006VariationalBF}
Ji, S., Krishnapuram, B. and Carin, L. (2006) Variational {B}ayes for
  continuous hidden {M}arkov models and its application to active learning.
\newblock \textit{IEEE Transactions on Pattern Analysis and Machine
  Intelligence}, \textbf{28}, 522--532.

\bibitem[{Johnson and Willsky(2014)}]{JohnsonWillsky2014}
Johnson, M.~J. and Willsky, A.~S. (2014) Stochastic variational inference for
  {B}ayesian time series models.
\newblock In \textit{ICML}, 1854--1862.

\bibitem[{Jordan et~al.(1999)Jordan, Ghahramani, Jaakkola and
  Saul}]{jordanetal1999}
Jordan, M.~I., Ghahramani, Z., Jaakkola, T.~S. and Saul, L.~K. (1999) An
  introduction to variational methods for graphical models.
\newblock \textit{Mach. Learn.}, \textbf{37}, 183–233.

\bibitem[{Kirshner(2005)}]{KirshnerThesis}
Kirshner, S. (2005) Modeling of multivariate time series using hidden {M}arkov
  models.
\newblock Ph.D.~Thesis, University of California, Irvine.

\bibitem[{Kullback and Leibler(1951)}]{KullbackLeibler1951}
Kullback, S. and Leibler, R.~A. (1951) On information and sufficiency.
\newblock \textit{Ann. Math. Stat.}, \textbf{22}, 79--86.

\bibitem[{MacKay(1997)}]{MacKay97ensemblelearning}
MacKay, D. J.~C. (1997) Ensemble learning for hidden {M}arkov models.
\newblock \textit{Tech. rep.}, Department of Physics, University of Cambridge.

\bibitem[{Majumder(2021)}]{ReetamThesis}
Majumder, R. (2021) Hidden {M}arkov models for high dimensional data with
  geostatistical applications.
\newblock Ph.D.~Thesis, University of Maryland, Baltimore County.

\bibitem[{Majumder et~al.(2020)Majumder, Mehta and Neerchal}]{MajumderJSM2020}
Majumder, R., Mehta, A. and Neerchal, N.~K. (2020) Copula-based correlation
  structure for multivariate emission distributions in hidden {M}arkov models.
\newblock In \textit{JSM Proceedings, Section on Statistics and the
  Environment}. VA: American Statistical Association.

\bibitem[{McGrory and Titterington(2009)}]{mcgroryetal}
McGrory, C.~A. and Titterington, D.~M. (2009) Variational {B}ayesian analysis
  for hidden {M}arkov models.
\newblock \textit{Aust. NZ J. Stat.}, \textbf{51}, 227--244.

\bibitem[{Mhanna and Bauwens(2012)}]{MhannaBauwens2012}
Mhanna, M. and Bauwens, W. (2012) A stochastic space-time model for the
  generation of daily rainfall in the {G}aza {S}trip.
\newblock \textit{Int.\ J. Climatol.}, \textbf{32}, 1098--1112.

\bibitem[{Rabiner(1989)}]{Rabiner1989ATO}
Rabiner, L.~R. (1989) A tutorial on hidden {M}arkov models and selected
  applications in speech recognition.
\newblock \textit{Proc. IEEE}, \textbf{77}, 257--286.

\bibitem[{Robbins and Monro(1951)}]{RobbinsMonro1951}
Robbins, H. and Monro, S. (1951) A stochastic approximation method.
\newblock \textit{Ann. Math. Stat.}, \textbf{22}, 400--407.

\bibitem[{Robertson et~al.(2004)Robertson, Kirshner and
  Smyth}]{RobertsonKirshnerSmyth2004}
Robertson, A.~W., Kirshner, S. and Smyth, P. (2004) Downscaling of daily
  rainfall occurrence over {N}ortheast {B}razil using a hidden {M}arkov model.
\newblock \textit{J. Climate}, \textbf{17}, 4407--4424.

\bibitem[{Robertson et~al.(2006)Robertson, Kirshner, Smyth, Charles and
  Bates}]{RobertsonEtAl2006}
Robertson, A.~W., Kirshner, S., Smyth, P., Charles, S.~P. and Bates, B.~C.
  (2006) Subseasonal-to-interdecadal variability of the {A}ustralian monsoon
  over {N}orth {Q}ueensland.
\newblock \textit{Q. J. Roy.\ Meteor.\ Soc.}, \textbf{132}, 519--542.

\bibitem[{Scott(2002)}]{scott2002}
Scott, S.~L. (2002) Bayesian methods for hidden {M}arkov models.
\newblock \textit{J. Am.\ Stat.\ Assoc.}, \textbf{97}, 337--351.

\bibitem[{{Viterbi}(1967)}]{Viterbi}
{Viterbi}, A. (1967) Error bounds for convolutional codes and an asymptotically
  optimum decoding algorithm.
\newblock \textit{IEEE T. Inform.\ Theory}, \textbf{13}, 260--269.

\bibitem[{Wilks(1998)}]{Wilks1998}
Wilks, D.~S. (1998) Multisite generalization of a daily stochastic
  precipitation generation model.
\newblock \textit{J. Hydrol.}, \textbf{210}, 178--191.

\end{thebibliography}
